\documentclass[%
12pt,
oneocolumn,
nofootinbib,
amsmath,amssymb,
aps,
floatfix,
unsortedaddress
]{revtex4-2}

\usepackage{graphicx,color}
\usepackage{subfig}
\usepackage{dcolumn}
\usepackage{bm}
\usepackage{float}
\usepackage{multirow}
\usepackage{mathtools}
\usepackage{cancel}
\usepackage{amsmath,amssymb,bm}
\usepackage{booktabs}
\usepackage{siunitx}
\usepackage{xcolor}

\usepackage[pdfpagelabels, pdfencoding=auto, psdextra]{hyperref}
\hypersetup{%
	pdfsubject=Paper,
	pdfkeywords={Interacting Boson Model, quantum phase transition,
		variational quantum eigensolver, quantum simulation, nuclear structure}
	unicode = true,
	breaklinks = true,
	colorlinks = true,
	linkcolor = blue,
	citecolor = blue,
	menucolor = blue,
	citecolor = blue,
	urlcolor = blue
}

\graphicspath{{./}}

\begin{document}
	\title{Symmetry-Reduced Variational Quantum Simulation of the $U(5)\rightarrow SU(3)$ Quantum Phase Transition in the Interacting Boson Model}
\author{Faisal Etminan}
\email{fetminan@birjand.ac.ir}
\affiliation{
	Department of Physics, Faculty of Sciences, University of Birjand, Birjand 97175-615, Iran
}%
\affiliation{ Interdisciplinary Theoretical and Mathematical Sciences Program (iTHEMS), RIKEN, Wako 351-0198, Japan}

\date{\today}%
	\begin{abstract}
The spherical-to-deformed quantum phase transition of the Interacting Boson Model (IBM) is investigated using the variational quantum eigensolver (VQE). The $U(5)$--$SU(3)$ transitional Hamiltonian is studied with $\chi=-\sqrt{7}/2$.
A symmetry-preserving, minimum-qubit VQE framework for collective nuclear models is developed, achieving a substantial reduction in qubit requirements without compromising the finite-size quantum-phase-transition physics.
The transition is characterized through the normalized $d$-boson occupation and ground-state energy derivatives. Finite-size results are found to approach the analytic critical point $\xi_c=8/17\simeq0.470588$, with an independent order-parameter extrapolation yielding $\xi_\infty=0.46986(61)$. The VQE reproduces ground-state energies and structural observables to numerical precision. These results demonstrate the potential of symmetry-reduced VQE for efficient quantum simulations of collective nuclear dynamics and quantum phase transitions.
	\end{abstract}
	
	\maketitle
	
	\section{Introduction}
	\label{sec:introduction}
	
	Quantum simulation provides a promising route toward studying
	many-body systems whose Hilbert spaces become increasingly difficult
	to treat with conventional computational methods~\cite{PhysRevLett.120.210501,PRXQuantum.5.020315, PhysRevD.109.114510}. Nuclear structure is
	a particularly interesting application because collective phenomena
	emerge from strongly correlated many-body degrees of freedom, while
	the relevant Hilbert spaces grow rapidly with particle number. Quantum
	computing methods, and in particular variational quantum algorithms,
	provide an alternative framework in which the ground-state problem can
	be formulated as an optimization of an expectation value on a quantum
	processor~\cite{RomeroPRC2022,KissPRC2022,StetcuPRC2022, RobinPRC2023, SinghPRC2025, HobdayPRC2025,Carrasco-CodinaPRC2026,GuPRC2026}.
	
	The Interacting Boson Model (IBM) provides a useful intermediate framework for investigating this possibility. Recent studies have explored the microscopic determination of IBM parameters using machine-learning methods for nuclear spectroscopy~\cite{OBATA2026140522}, as well as the use of entanglement entropy as a signal of quantum phase transitions in even-even and odd-$A$ nuclei~\cite{JafarizadehPRC2022}.
		This model introduced by Arima and Iachello~\cite{Iachello_Arima_1987} is
	an algebraic model that has its roots in the nuclear shell model.
	In IBM-1, low-lying
	collective states of even-even nuclei are described in terms of
	correlated pairs of valence nucleons represented by $s$ and $d$ bosons.
	Despite its relatively compact formulation, the IBM contains the
	essential ingredients required for studying collective shape evolution
	and quantum phase transitions. In particular, the consistent-$Q$
	Hamiltonian provides a continuous path between the spherical $U(5)$
	and deformed $SU(3)$ dynamical-symmetry limits.
	
	The $U(5)\rightarrow SU(3)$ transition is therefore a natural
	benchmark problem for quantum simulation. It contains a well-defined
	thermodynamic-limit critical point, while finite boson number produces
	a smooth crossover that can be studied numerically. 
		The concept of quantum phase transitions was introduced by Gilmore \textit{et al.}~\cite{GILMORE197826,PhysRevC.19.1119} in analogy with conventional thermodynamic phase transitions. Unlike thermodynamic phase transitions, which are driven by temperature, quantum phase transitions arise from variations in the parameters of the Hamiltonian governing the quantum system.

	The IBM also makes
	it possible to separate several aspects of the quantum-simulation
	problem: construction of the many-body Hamiltonian, symmetry
	reduction, qubit encoding, variational optimization, and identification
	of the phase-transition signatures.
	
	The purpose of the present work is to investigate this complete chain
	systematically. 
	The IBM Hamiltonian is first constructed directly in the occupation-number basis, and the resulting matrix representation is validated.
	The exact ground-state properties are then investigated as functions
	of the boson number and the control parameter. Finite-size scaling is
	used to establish the convergence toward the known thermodynamic-limit
	critical point.
	
	A central part of the present study is the use of rotational symmetry
	to reduce the Hilbert space before quantum encoding. Since the
	Hamiltonian is rotationally invariant and the ground state belongs to
	the $L=0$ sector, the physically relevant Hilbert space can be much
	smaller than the complete fixed-$N_B$ IBM space. This reduction is
	followed by a minimum-qubit binary encoding and an explicit validation
	against the full IBM Hamiltonian.
	
	The resulting qubit Hamiltonian is subsequently treated with VQE.
	Rather than restricting the comparison to the ground-state energy, structural observables and quantum-phase-transition indicators are also calculated from the variational state. Finally, a thermodynamic-limit analysis is performed using independent energy- and order-parameter-based criteria, and their extrapolated critical points are compared.

	The paper is organized as follows. Section~\ref{sec:theoretical_framework} introduces the IBM Hamiltonian and its dynamical-symmetry limits. Section~\ref{sec:ground_state_finite_size} presents the exact ground-state properties and finite-size analysis of the quantum phase transition. Section~\ref{sec:symmetry_encoding} discusses the $L=0$ symmetry reduction and minimum-qubit encoding. Section~\ref{sec:VQE} introduces the variational quantum eigensolver (VQE) implementation and its validation. Section~\ref{sec:VQE_QPT} analyzes the quantum phase transition using VQE, while Section~\ref{sec:thermodynamic_limit} presents the thermodynamic-limit extrapolation. The results and their physical implications are discussed in Sec.~\ref{sec:discussion}. Finally, conclusions are given in Sec.~\ref{sec:conclusions}.
	
	\section{Theoretical framework}
	\label{sec:theoretical_framework}
	
	\paragraph*{Interacting Boson Model.}
	\label{subsec:IBM}
	
	The Interacting Boson Model (IBM) provides a phenomenological
	description of the low-lying collective states of even-even nuclei in
	terms of correlated pairs of valence nucleons. In the simplest IBM-1
	formulation, the relevant degrees of freedom are represented by one
	$s$ boson with angular momentum $L=0$ and five $d$ bosons with
	angular momentum $L=2$. The corresponding creation operators are
	denoted by $s^\dagger$ and $d^\dagger_\mu$, with
	$\mu=-2,-1,0,1,2$.
	
	The model therefore contains six bosonic modes, $ s, d_{-2},d_{-1},d_0,d_1,d_2 $. 	
	The total number of bosons is conserved,
	\begin{equation}
		\hat N_B
		=
		s^\dagger s+
		\sum_{\mu=-2}^{2}d^\dagger_\mu d_\mu ,
	\end{equation}
	and calculations are performed in a subspace with fixed boson number, $\hat N_B=N_B$.	
	A convenient occupation-number basis is $ |n_s,n_{-2},n_{-1},n_0,n_1,n_2\rangle	$,
	subject to 	$ n_s+\sum_{\mu=-2}^{2}n_\mu=N_B $.

	Since $N_B$ identical bosons are distributed among six bosonic modes, the dimension of the complete symmetric IBM-1 Hilbert space is
	\begin{equation}
				D(N_B) = \binom{N_B+5}{5} = \frac{(N_B+1)(N_B+2)(N_B+3)(N_B+4)(N_B+5)}{120}.
		\label{eq:IBM_dimension}
	\end{equation}
	This observation is relevant to the computational analysis developed below: although the fixed-$N_B$ IBM Hilbert-space dimension is polynomial in $N_B$, its rapid increase with boson number leads to increasing computational costs for classical many-body calculations and motivates the investigation of alternative quantum-computing strategies.
	
	The bosonic creation and annihilation operators satisfy the canonical commutation relations $	[b_i,b_j^\dagger]=\delta_{ij}$, where $i$ and $j$ label the six bosonic modes. Their action on an occupation-number state is given by
	\begin{align}
	b_i^\dagger|\ldots,n_i,\ldots\rangle
	&=
	\sqrt{n_i+1}
	|\ldots,n_i+1,\ldots\rangle,
	\\
	b_i|\ldots,n_i,\ldots\rangle
	&=
	\sqrt{n_i}
	|\ldots,n_i-1,\ldots\rangle.
    \end{align}
	These relations provide the basis for the explicit construction of the IBM operators in the occupation-number representation.
	\paragraph*{The IBM Hamiltonian.}
	\label{subsec:IBM_Hamiltonian}
	
	The most compact form of the IBM Hamiltonian that captures the essential physics of the model is employed, as obtained within the consistent-$Q$ formalism~\cite{WarnerPhysRevC1983,HELLEMANS2007164}
		\begin{equation}
			\hat H(\xi,\chi)
			=
			(1-\xi)\hat n_d
			-
			\frac{\xi}{4N_B}
			\hat Q_\chi\cdot\hat Q_\chi ,
		\label{eq:IBM_Hamiltonian}
	\end{equation}
	the two terms in Eq.~\eqref{eq:IBM_Hamiltonian} represent competing structural tendencies. The first term favors configurations with a small number of $d$ bosons, whereas the attractive quadrupole-quadrupole interaction in the second term favors collective quadrupole correlations. Varying $0\leq\xi\leq1$ therefore drives the system from a predominantly spherical configuration toward a quadrupole-deformed configuration.	
	The $d$-boson number operator is
	\begin{equation}
		\hat n_d
		=
		\sum_{\mu=-2}^{2}
		d^\dagger_\mu d_\mu .
		\label{eq:nd}
	\end{equation} 	
	The quadrupole operator is
	\begin{equation}
		\hat Q_{\chi,\mu}
		=
		d^\dagger_\mu s
		+
		s^\dagger\tilde d_\mu
		+
		\chi
		[d^\dagger\times\tilde d]^{(2)}_\mu ,
		\label{eq:quadrupole_operator}
	\end{equation}
	where $	\tilde d_\mu=(-1)^\mu d_{-\mu}$.	
	The tensor-coupled term is
	\begin{equation}
		[d^\dagger\times\tilde d]^{(2)}_\mu
		=
		\sum_{m,m'}
		\langle2m,2m'|2\mu\rangle
		d^\dagger_m\tilde d_{m'} .
		\label{eq:d_tensor}
	\end{equation} 	
	The scalar product is defined by
	\begin{equation}
		\hat Q_\chi\cdot\hat Q_\chi
		=
		\sum_{\mu=-2}^{2}
		(-1)^\mu
		\hat Q_{\chi,\mu}\hat Q_{\chi,-\mu}.
		\label{eq:QQ}
	\end{equation}
		Both $\hat n_d$ and $\hat Q_\chi$ conserve the total boson number.
	Therefore, $ [\hat H,\hat N_B]=0 $,	and the Hamiltonian acts entirely within the fixed-$N_B$ Hilbert
	space.
	
	\paragraph*{Dynamical symmetry limits.}
	\label{subsec:symmetry_limits}
	
	At $\xi=0$, the Hamiltonian becomes $ \hat H=\hat n_d$, corresponding to the spherical vibrational $U(5)$ limit. 
	For the prolate $SU(3)$ limit, the following choice is made:
	\begin{equation}
		\chi=-\frac{\sqrt7}{2},
		\qquad
		\xi=1.
		\label{eq:SU3_limit}
	\end{equation}	
	The parameter $\xi$ therefore controls the structural evolution,
	whereas $\chi$ determines the geometric character of the quadrupole
	operator. 
	Throughout the principal transition studied in this work, $\chi=-{\sqrt7}/{2}$ is fixed, while $\xi$ is varied between zero and unity.
	
	\paragraph*{Intrinsic coherent state and nuclear shape.}
	\label{subsec:coherent_state}
	Despite its microscopic foundation in the nuclear shell model, the IBM can also be connected to a macroscopic, geometrical description of the nucleus through the coherent-state formalism~\cite{GinocchioPhysRevLett1980,DieperinkPhysRevLett1980,AageBohr1980},
	\begin{equation}
		|\beta,\gamma;N_B\rangle
		=
		\frac{1}{\sqrt{N_B!}}
		(b_c^\dagger)^{N_B}|0\rangle ,
		\label{eq:coherent_state}
	\end{equation}
	where
	\begin{equation}
		b_c^\dagger
		=
		\frac{1}{\sqrt{1+\beta^2}}
		\left[
		s^\dagger
		+\beta\cos\gamma\,d_0^\dagger
		+\frac{\beta}{\sqrt2}\sin\gamma
		(d_2^\dagger+d_{-2}^\dagger)
		\right].
		\label{eq:coherent_boson}
	\end{equation}	
	The variable $\beta$ measures the magnitude of the intrinsic
	quadrupole deformation, while $\gamma$ characterizes triaxiality.
	The spherical configuration corresponds to $\beta=0$.	
	The expectation value of the $d$-boson number is
	\begin{equation}
		\langle\hat n_d\rangle
		=
		N_B\frac{\beta^2}{1+\beta^2},
		\label{eq:nd_coherent}
	\end{equation}
	and hence
	\begin{equation}
		\frac{\langle\hat n_d\rangle}{N_B}
		=
		\frac{\beta^2}{1+\beta^2}.
		\label{eq:nd_fraction_beta}
	\end{equation}
	
	\paragraph*{Finite-$N_B$ coherent-state energy surface.}
	\label{subsec:finite_coherent_surface}
	
	For the normalized coherent state, the expectation value of the
	quadrupole interaction is
	\begin{align}
		\langle
		\hat Q_\chi\cdot\hat Q_\chi
		\rangle
		={}&
		N_B
		\frac{5+(1+\chi^2)\beta^2}
		{1+\beta^2}
		\nonumber\\
		&+
		N_B(N_B-1)
		\frac{
			4\beta^2
			-4\sqrt{\frac27}\chi\beta^3\cos3\gamma
			+\frac27\chi^2\beta^4
		}
		{(1+\beta^2)^2}.
		\label{eq:QQ_coherent}
	\end{align}
	
	The finite-$N_B$ intrinsic energy surface is therefore~\cite{HELLEMANS2007164},
	\begin{align}
		E_{\rm coh}(\beta,\gamma)
		={}&
		(1-\xi)N_B
		\frac{\beta^2}{1+\beta^2}
		\nonumber\\
		&-
		\frac{\xi}{4}
		\frac{5+(1+\chi^2)\beta^2}
		{1+\beta^2}
		\nonumber\\
		&-
		\frac{\xi(N_B-1)}{4}
		\frac{
			4\beta^2
			-4\sqrt{\frac27}\chi\beta^3\cos3\gamma
			+\frac27\chi^2\beta^4
		}
		{(1+\beta^2)^2}.
		\label{eq:energy_surface_exact}
	\end{align}	
	Introducing the energy per boson, $ \mathcal E_{\rm coh} = {E_{\rm coh}}/{N_B}$, the thermodynamic-limit surface becomes
	\begin{align}
		\mathcal E_\infty(\beta,\gamma)
		={}&
		(1-\xi)
		\frac{\beta^2}{1+\beta^2}
		\nonumber\\
		&-
		\frac{\xi}{4}
		\frac{
			4\beta^2
			-4\sqrt{\frac27}\chi\beta^3\cos3\gamma
			+\frac27\chi^2\beta^4
		}
		{(1+\beta^2)^2}.
		\label{eq:energy_surface_largeN}
	\end{align}	
	For the $U(5)\rightarrow SU(3)$ transitional path, minimization of the intrinsic energy surface yields the critical value
	\begin{equation}
					\xi_c=\frac{14}{28+\chi^2}.
				\label{eq:critical_general}
	\end{equation}
	For $\chi=-\sqrt7/2$, this reduces to $ \xi_c={8}/{17} 	\simeq0.470588$.
	At the critical point, the spherical and deformed minima coexist. The deformed minimum is located at
	\begin{equation}
		\beta_c=\frac{1}{2\sqrt2},
		\qquad
		\gamma_c=0.
		\label{eq:critical_beta}
	\end{equation}
	This coexistence is characteristic of the first-order nature of the
	thermodynamic-limit $U(5)\rightarrow SU(3)$ transition.

	Although the critical point $\xi_c=8/17$ is obtained in the thermodynamic limit from the mean-field energy surface, the finite-$N_B$ behavior of the IBM can differ appreciably from this limiting result. Finite-size effects in IBM quantum phase transitions have been studied systematically beyond the mean-field approximation. In particular, finite-size scaling at the $U(5)\rightarrow O(6)$ critical point has been investigated using a Holstein--Primakoff expansion combined with continuous unitary transformations~\cite{Vidal-PhysRevC.2005}. Related studies by Vidal \textit{et al.} introduced a scalar two-level boson model with the same energy surface as the $Q$-consistent IBM Hamiltonian and used it to analyze finite-size corrections to the energy gap and order parameter near criticality~\cite{Vidal-PhysRevC.2006}. The limitations of the standard intrinsic-state, or mean-field, description for finite systems have also been examined using two-level boson models beyond mean field, with analytical results compared with exact diagonalization~\cite{Vidal-PhysRevC.2007}.
	
	These studies motivate an explicit finite-$N_B$ analysis in the present work. Rather than identifying the finite-system transition directly with the thermodynamic-limit value $\xi_c$, the transition is characterized through ground-state observables, a finite-$N_B$ pseudocritical point $\xi_{\rm pc}(N_B)$ is determined, and its systematic evolution with boson number is investigated. In this way, a quantitative connection is established between the finite systems accessible to exact diagonalization and quantum simulation and the thermodynamic-limit critical point of the IBM.
		
\section{Exact IBM ground-state properties and finite-size analysis}
\label{sec:ground_state_finite_size}

Having established the IBM Hamiltonian and its operator representation,
The ground-state properties are first investigated using exact diagonalization. These calculations provide the classical reference against which all subsequent quantum-computing results are compared. Throughout this section, $\chi=-{\sqrt{7}}/{2}$ is fixed, while the control parameter is varied over the range $0\leq\xi\leq1$.

For each value of $N_B$ and $\xi$, the Hamiltonian is diagonalized in the
complete fixed-$N_B$ IBM Hilbert space.  The ground state is obtained from
\begin{equation}
	\hat H(\xi,\chi)|\Psi_0(\xi,\chi)\rangle
	=
	E_0(\xi,\chi)|\Psi_0(\xi,\chi)\rangle.
	\label{eq:ground_state_exact}
\end{equation}
The principal quantities used to characterize the structural evolution
are the ground-state energy,
the normalized $d$-boson occupation $\left(\rho_{d}\right)$,
and the quadrupole collectivity $ \left(C_Q\right)$,
\begin{equation}
		\rho_d(\xi)
	=
	\frac{\langle\Psi_0|\hat n_d|\Psi_0\rangle}{N_B},
	\qquad
	C_Q(\xi)
	=
	\frac{\langle\Psi_0|
		\hat Q_\chi\cdot\hat Q_\chi
		|\Psi_0\rangle}{N_B^2}.
	\label{eq:CQ_main}
\end{equation}
The quantity $\rho_d$ is particularly useful because it provides a
direct measure of the redistribution of bosons from the spherical
$s$-boson configuration toward configurations containing quadrupole
$d$ bosons.  It therefore serves as the primary order parameter in the
finite-$N_B$ analysis.
	
	\paragraph*{Ground-state energy and order parameter.}
	
	For small $\xi$, $\rho_d$ is close to zero, reflecting the spherical
	$U(5)$ structure. Increasing $\xi$ strengthens quadrupole correlations
	and produces a rapid increase of $\rho_d$ in the transition region.
	
\begin{figure}[t]
		\centering
		\includegraphics[width=0.72\textwidth]{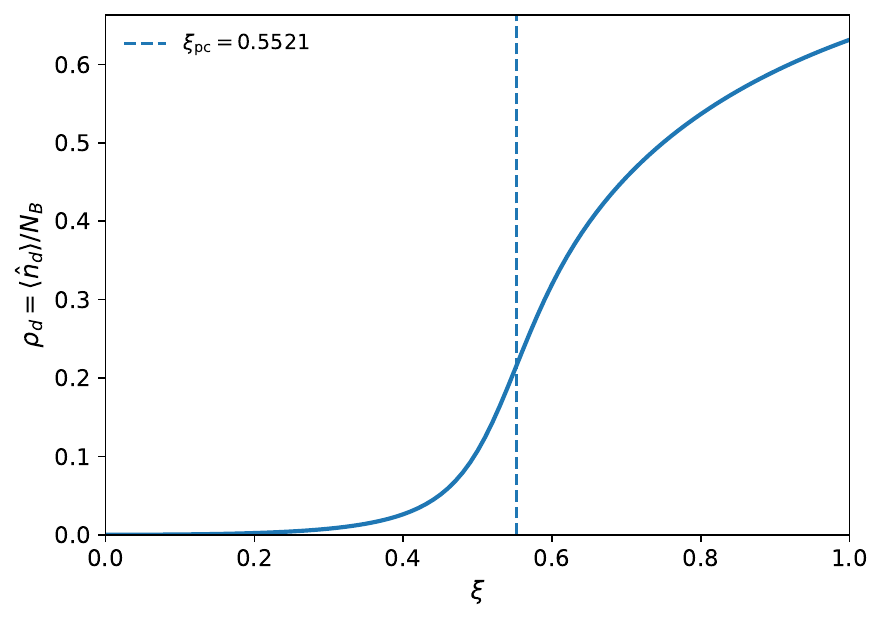}
		\caption{
			Normalized $d$-boson occupation
			$\rho_d=\langle\hat n_d\rangle/N_B$ as a function of the control
			parameter $\xi$ for $N_B=10$ and
			$\chi=-\sqrt{7}/2$.  The rapid increase in the transitional region
			signals the evolution from the spherical $U(5)$ regime toward the
			deformed $SU(3)$ regime.  The vertical line indicates the
			finite-$N_B$ pseudocritical point obtained from the maximum of
			$d\rho_d/d\xi$.
		}
		\label{fig:rho_d_NB10}
	\end{figure}

	Figure~\ref{fig:rho_d_NB10} displays the normalized $d$-boson occupation
	for $N_B=10$ over the full range of the control parameter.  At small
	$\xi$, $\rho_d$ remains close to zero, reflecting the predominantly
	spherical $U(5)$ structure.  With increasing $\xi$, the order parameter
	undergoes a pronounced increase in the transitional region and approaches
	a substantially larger value on the deformed side of the phase diagram.
	The transition is nevertheless smooth for finite $N_B$, as expected for
	a finite quantum system.
	
	This behavior provides the first numerical indication of the
	$U(5)\rightarrow SU(3)$ structural evolution.  More importantly, the
	rapid variation of $\rho_d$ suggests that its derivative can be used to
	define a finite-size pseudocritical point.

	For a finite boson number the transition does not produce a
	nonanalyticity.  The pseudocritical point is therefore defined by the maximum response of the order parameter,
	\begin{equation}
		\xi_{\rm pc}(N_B)
		=
		\operatorname*{arg\,max}_{\xi}
		\left[
		\frac{d\rho_d}{d\xi}
		\right],
		\label{eq:xipc_definition_final}
	\end{equation}
	with
	\begin{equation}
		D_{\max}(N_B)
		=
		\max_{\xi}
		\left[
		\frac{d\rho_d}{d\xi}
		\right].
		\label{eq:Dmax_definition_final}
	\end{equation}
	
	The calculated pseudocritical points systematically decrease as the
	boson number increases.  This finite-size evolution is summarized in
	Table~\ref{tab:finite_size_results}.  The corresponding increase of
	$D_{\max}$ demonstrates that the crossover becomes progressively sharper
	with increasing $N_B$.

	\paragraph*{Finite-$N_B$ pseudocritical point.}
	
	The calculated finite-size pseudocritical points are summarized in
	Table~\ref{tab:finite_size_results}.
	\begin{table}[ht]
		\centering
		\caption{Finite-$N_B$ pseudocritical points and maximum slopes
			obtained for $\chi=-\sqrt7/2$.}
		\label{tab:finite_size_results}
		\begin{tabular}{cccc}
			\toprule
			$N_B$ & $D_{\rm IBM}$ & $\xi_{\rm pc}$ & $D_{\max}$\\
			\midrule
			6  & 462   & 0.5974 & 1.773998\\
			8  & 1287  & 0.5696 & 2.082825\\
			10 & 3003  & 0.5521 & 2.352399\\
			12 & 6188  & 0.5397 & 2.595446\\
			14 & 11628 & 0.5305 & 2.819558\\
			16 & 20349 & 0.5234 & 3.029648\\
			18 & 33649 & 0.5176 & 3.229096\\
			20 & 53130 & 0.5129 & 3.420322\\
			\bottomrule
		\end{tabular}
	\end{table}
	
	The pseudocritical point decreases systematically with increasing
	boson number, demonstrating convergence toward the thermodynamic-limit
	critical region.
	
	\paragraph*{Finite-size scaling.}
	
	The finite-size shift may be parameterized as
	\begin{equation}
		\Delta\xi(N_B)
		=
		\xi_{\rm pc}(N_B)-\xi_c.
	\end{equation}	
	A power-law fit,
	\begin{equation}
		\Delta\xi(N_B)=AN_B^{-\alpha},
	\end{equation}
to the results for $6\leq N_B\leq20$ gives $ A=0.6361(13)$, $\alpha=0.8968(88)$, with $	R^2=0.99943$.
The excellent quality of the fit indicates that the finite-size shift
is well described by a power law over the investigated range. However,
the fitted exponent exhibits a gradual dependence on the fitting range,
suggesting that the pure power-law form should be regarded as an
effective finite-size parametrization rather than as an asymptotic
scaling law.

	An inverse-boson-number expansion,
	\begin{equation}
		\Delta\xi(N_B)
		=
		\frac{A}{N_B}
		+
		\frac{B}{N_B^2},
	\end{equation}
	gives $ A=0.8931(26)$ and $B=-0.7942(20)$ with $R^2=0.99995$.	The corresponding residual sum of squares is $	{\rm RSS}=2.99\times10^{-7}$,
	which is substantially smaller than that obtained from the leading
	$1/N_B$ form alone. This indicates that the subleading $1/N_B^2$
	correction provides an important improvement for the finite boson
	numbers considered here.

\begin{figure}[t]
		\centering
		\includegraphics[width=0.72\textwidth]{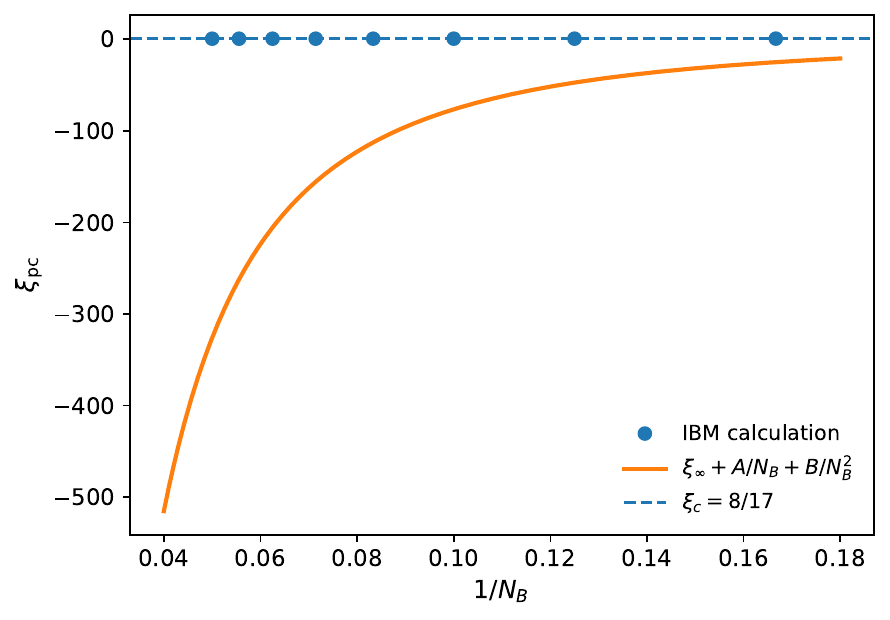}
		\caption{
			Finite-size evolution of the pseudocritical point
			$\xi_{\rm pc}$ as a function of inverse boson number $1/N_B$.
			The points are obtained from the maximum of
			$d\rho_d/d\xi$, while the solid curve represents the fit
			$\xi_{\rm pc}=\xi_\infty+A/N_B+B/N_B^2$.
			The horizontal line indicates the analytical large-$N_B$
			critical point $\xi_c=8/17$.
		}
		\label{fig:finite_size_xipc}
	\end{figure}

	The convergence of the pseudocritical point can be examined by plotting
	$\xi_{\rm pc}$ against the inverse boson number.  As shown in
	Fig.~\ref{fig:finite_size_xipc}, the calculated points move
	systematically toward the large-$N_B$ critical region.  
	This behavior is described using the expansion
	\begin{equation}
		\xi_{\rm pc}(N_B)
		=
		\xi_\infty
		+
		\frac{A}{N_B}
		+
		\frac{B}{N_B^2}.
		\label{eq:xipc_free_fit_final}
	\end{equation}	
	The extrapolated value, $	\xi_\infty=0.46986(61)$, with $A=0.9081(13)$, $B=-0.8613(60)$ and $R^2=0.99996$ 
	is in excellent agreement with the analytical critical point
	$\xi_c=8/17\simeq0.470588$. The difference from the analytical result is $\xi_\infty-\xi_c = -7.25\times10^{-4}$, 
	demonstrating excellent agreement.

\paragraph*{Finite-size sharpening of the transition.}
	
In addition to the movement of the pseudocritical point, the crossover
becomes increasingly sharp with increasing boson number. This behavior
is quantified by the maximum slope $D_{\max}$ defined in
Eq.~\eqref{eq:Dmax_definition_final}.

The calculated values of $D_{\max}$ increase systematically from
$1.774$ for $N_B=6$ to $3.420$ for $N_B=20$. This dependence is described using the finite-size scaling form 
\begin{equation}
	D_{\max}(N_B)
	=
	C N_B^\gamma.
	\label{eq:Dmax_power}
\end{equation}
A fit over the range $6\leq N_B\leq20$ gives $C=0.67329(23)$ and $\gamma=0.54260(13)$ with
$R^2=0.99997$. The positive value of $\gamma$ demonstrates that the maximum response
of the order parameter increases with the number of bosons. Thus, the
smooth crossover characteristic of finite systems becomes progressively
sharper as the large-$N_B$ limit is approached.

\begin{figure}[t]
		\centering
		\includegraphics[width=0.72\textwidth]{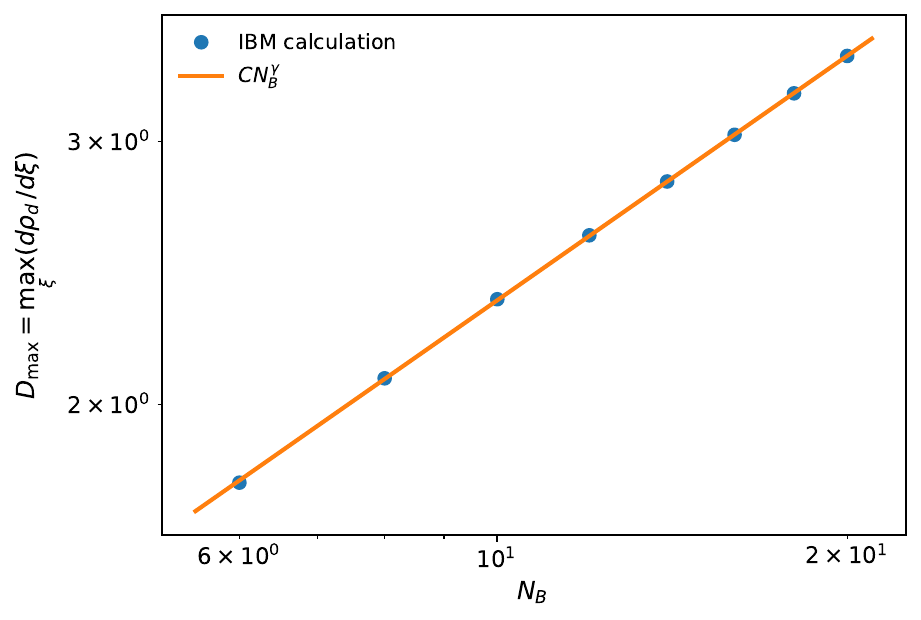}
		\caption{
			Maximum response of the order parameter,
			$D_{\max}=\max_\xi[d\rho_d/d\xi]$, as a function of the boson
			number $N_B$.  The solid curve represents the power-law fit
			$D_{\max}=C N_B^\gamma$.  The systematic increase of
			$D_{\max}$ demonstrates the progressive sharpening of the
			finite-$N_B$ crossover.
		}
		\label{fig:slope_scaling}
	\end{figure}

The exponent $\gamma$ is regarded here as a finite-size scaling exponent
characterizing the numerical growth of the maximum slope. It is not
interpreted as a universal critical exponent, since establishing
universality would require a dedicated finite-size scaling study over a
substantially wider range of boson numbers and a more detailed analysis
of the large-$N_B$ limit. The resulting finite-size scaling of the
maximum slope is shown in Fig.~\ref{fig:slope_scaling}.
	
\paragraph*{Summary of the exact finite-size analysis.}
	
The finite-$N_B$ calculations provide two complementary signatures of
the $U(5)$--$SU(3)$ quantum phase transition. First, the pseudocritical
point extracted from the maximum derivative of the order parameter
moves systematically toward the known large-$N_B$ critical point,
$\xi_c=8/17$. An independent extrapolation of the finite-$N_B$
pseudocritical points gives $\xi_\infty=0.46986(61)$, consistent with
the exact value within the numerical uncertainty. Second, the maximum
slope of the order parameter increases approximately as
$N_B^{0.5426}$, demonstrating the progressive sharpening of the
transition with increasing boson number.

These results establish the classical IBM ground-state behavior that
will serve as a benchmark for the quantum simulation.
 In the following section, the ability of the variational quantum eigensolver to reproduce the same ground-state properties and the associated quantum phase transition is therefore investigated.

\section{Symmetry reduction and quantum encoding}
\label{sec:symmetry_encoding}

The complete fixed-$N_B$ IBM Hilbert space provides the natural
reference space for exact diagonalization.  For quantum simulation,
however, this space contains many states that are not relevant to the
ground state considered in the present work.  The IBM Hamiltonian is
rotationally invariant and therefore conserves total angular momentum.
Since the ground state studied here belongs to the $L=0$ sector, the
rotational symmetry can be exploited before the qubit encoding and
variational optimization are performed.

This symmetry reduction is particularly important for the present
study because it provides not only a reduction in the number of qubits,
but also a physically motivated variational space for the VQE.  The
resulting symmetry-aware ansatz is therefore constructed entirely
within the relevant $L=0$ sector.

\paragraph*{Rotational symmetry and projection onto the $L=0$ sector.}
\label{subsec:L0_projection}

For a fixed total number of bosons $N_B$, the complete IBM Hilbert-space
dimension is given by Eq.~\eqref{eq:IBM_dimension}. The IBM angular-momentum operator is constructed from the $d$-boson
operators as
\begin{equation}
	\hat L_\mu
	=
	\sqrt{10}
	\left[d^\dagger\times\tilde d\right]^{(1)}_\mu ,
	\qquad
	\mu=-1,0,1.
	\label{eq:L_operator}
\end{equation}

The total angular momentum operator is $\hat{L}^{2}=\sum_{\mu=-1}^{1}(-1)^{\mu}\hat{L}_{\mu}\hat{L}_{-\mu}$ .
The IBM Hamiltonian considered in this work is rotationally invariant
and consequently satisfies $ [\hat H,\hat L^2]=0$ and $[\hat H,\hat L_z]=0$.
The Hilbert space therefore decomposes into mutually orthogonal
angular-momentum sectors,
\begin{equation}
	\mathcal H_{\rm IBM}
	=
	\bigoplus_L\mathcal H_{N_B,L}.
	\label{eq:IBM_L_decomposition}
\end{equation}
The ground state considered in the present work has $L=0$.  The relevant
physical subspace is consequently $ \mathcal H_{N_B,L=0} \subset \mathcal H_{\rm IBM}$, with dimension
$ D_{L=0}(N_B) = \dim\mathcal H_{N_B,L=0}$.
Because the Hamiltonian preserves angular momentum, projection onto
this sector is an exact symmetry reduction rather than an
approximation.  States belonging to $L\neq0$ cannot mix with the
$L=0$ ground state under the action of $\hat H$.  The ground-state
problem can therefore be solved entirely within
$\mathcal H_{N_B,L=0}$.

An important distinction is that the reduction is performed with
respect to angular momentum and not with respect to the $d$-boson
number.  In particular, $[\hat H,\hat n_d]\neq0$,
because the quadrupole-quadrupole interaction couples different
$n_d$ configurations.  Consequently, all physically allowed
$n_d$ configurations contained in the $L=0$ sector are retained.
The symmetry reduction therefore removes only states forbidden by the
conserved angular momentum, while preserving the complete dynamical
mixing required by the IBM Hamiltonian.

\paragraph*{Dimension of the symmetry-reduced space.}
\label{subsec:L0_dimension}

The complete Hilbert-space dimension, $D_{\rm IBM}$ grows as shown in Eq.~\eqref{eq:IBM_dimension}, whereas the dimension relevant for the present ground-state problem is
only $D_{L=0} = \dim\mathcal H_{N_B,L=0}$. 
The dimensions obtained from the explicit construction of the IBM basis
and angular-momentum projection are listed in
Table~\ref{tab:qubit_resources}.  The corresponding minimum number of
qubits is also shown.

\begin{table}[ht]
	\centering
	\caption{Dimensions of the complete IBM Hilbert space, the
		symmetry-reduced $L=0$ sector, and the minimum number of qubits
		required for an exact binary encoding.}
	\label{tab:qubit_resources}
	\begin{tabular}{cccc}
		\toprule
		$N_B$ & $D_{\rm IBM}$ & $D_{L=0}$ & $N_q$\\
		\midrule
		2  & 21   & 2  & 1\\
		3  & 56   & 3  & 2\\
		4  & 126  & 4  & 2\\
		5  & 252  & 5  & 3\\
		6  & 462  & 7  & 3\\
		8  & 1287 & 10 & 4\\
		10 & 3003 & 14 & 4\\
		\bottomrule
	\end{tabular}
\end{table}

The resource reduction should be understood as a consequence of using
the physical symmetry sector before the quantum encoding, rather than
as an approximation to the IBM Hilbert space.

\paragraph*{Minimum-qubit encoding.}
\label{subsec:minimum_qubit_encoding}

Let $ \{|L=0,\alpha\rangle\}_{\alpha=0}^{D_{L=0}-1}$
denote an orthonormal basis of the physical $L=0$ sector.  These basis
states can be mapped one-to-one onto computational-basis states of a
qubit register.  The minimum number of qubits required to represent the
sector is
\begin{equation}
		N_q
		=
		\left\lceil
		\log_2D_{L=0}
		\right\rceil .
	\label{eq:Nq_minimum}
\end{equation}
When $D_{L=0}$ is not an exact power of two, the qubit Hilbert space
contains additional computational-basis states that do not correspond
to physical IBM states.  These states are identified as unphysical computational-basis configurations of the binary representation and are excluded
from the physical Hamiltonian construction.  Equivalently, defining
the physical projector
\begin{equation}
	\hat P_{\rm phys}
	=
	\sum_{\alpha=0}^{D_{L=0}-1}
	|\mathrm{bin}(\alpha)\rangle
	\langle\mathrm{bin}(\alpha)|,
	\label{eq:Pphys}
\end{equation}
the encoded Hamiltonian acts only within the physical subspace
selected by $\hat P_{\rm phys}$.

For comparison, a direct binary encoding of the complete IBM Hilbert
space would require
\begin{equation}
	N_q^{\rm full}
	=
	\left\lceil
	\log_2D_{\rm IBM}
	\right\rceil .
	\label{eq:Nq_full}
\end{equation}
Thus, the reduction from $D_{\rm IBM}$ to $D_{L=0}$ translates directly
into a reduction of the minimum qubit requirement.  For example, for $N_B=6$, a direct binary encoding of the complete
$D_{\rm IBM}=462$ dimensional IBM space would require
$N_q^{\rm full}=9$ qubits, whereas the symmetry-reduced
$D_{L=0}=7$ dimensional sector requires only $N_q=3$ qubits.
Thus, exploiting the rotational symmetry reduces the minimum qubit
requirement from nine to three for this case.
\paragraph*{Embedding and penalty for unused computational states.}
For the direct symmetry-reduced VQE used in the QPT calculations, no penalty term is required because the variational state is represented entirely within the physical $L=0$ space. The penalty construction is relevant only when the reduced Hamiltonian is embedded into the full minimum-qubit computational Hilbert space, as in the independent qubit-encoding and circuit benchmarks.

When $D_{L=0}$ is not a power of two, the minimum-qubit register
contains computational-basis states that have no counterpart in the
physical $L=0$ IBM sector. The physical Hamiltonian is therefore embedded as a block within the full qubit-space Hamiltonian.
 Specifically, if
$D_{L=0}<2^{N_q}$, the encoded Hamiltonian is constructed as
\begin{equation}
	\hat H_q
	=
	\hat H_{\rm phys}
	+
	\lambda \hat P_{\rm unphys},
	\label{eq:penalty_hamiltonian}
\end{equation}
where $\hat P_{\rm unphys} 	= \sum_{\alpha=D_{L=0}}^{2^{N_q}-1} |\alpha\rangle\langle\alpha|$
projects onto the unused computational-basis states and
$\lambda>0$ is an energy penalty. In the computational basis used in
the present implementation, the physical Hamiltonian occupies the
upper-left $D_{L=0}\times D_{L=0}$ block, while each unphysical basis
state is assigned the diagonal energy $\lambda$:
\begin{equation}
	H_q =
	\begin{pmatrix}
		H_{\rm phys} & 0\\
		0 & \lambda I_{2^{N_q}-D_{L=0}}
	\end{pmatrix}.
	\label{eq:embedded_hamiltonian}
\end{equation}
Consequently, the penalty does not alter any eigenvalue or eigenvector
within the physical subspace. Its purpose is to place the unused
computational states sufficiently high in energy that they cannot
become the variational ground state.

In the present implementation, when no penalty is specified
explicitly, its value is chosen automatically from a Gershgorin
upper bound on the magnitude of the IBM Hamiltonian eigenvalues,
\begin{equation}
	\lambda
	=
	10\left[
	\max_i\sum_j |(H_{\rm phys})_{ij}|+1
	\right].
	\label{eq:automatic_penalty}
\end{equation}
This choice provides a conservative energy separation between the
physical spectrum and the unused computational states. The resulting
VQE states are subsequently checked by evaluating their probability
within the physical subspace,
\begin{equation}
	P_{\rm phys}
	=
	\langle\psi|
	\hat P_{\rm phys}
	|\psi\rangle,
	\qquad
	\hat P_{\rm phys}=I-\hat P_{\rm unphys}.
	\label{eq:physical_probability}
\end{equation}
For all calculations reported below, $P_{\rm phys}$ is unity within
numerical precision, confirming that the optimized states remain in
the intended physical Hilbert space.

\paragraph*{Construction of the symmetry-reduced Hamiltonian.}
\label{subsec:encoded_hamiltonian}
  The Hamiltonian is projected onto orthonormal basis $ \{|L=0,\alpha\rangle\}_{\alpha=0}^{D_{L=0}-1}$,
\begin{equation}
	H^{(L=0)}_{\alpha\beta}
	=
	\langle L=0,\alpha|
	\hat H
	|L=0,\beta\rangle .
	\label{eq:H_L0}
\end{equation}
Since $ [\hat H,\hat L^2]=0$, the $L=0$ sector is invariant under the Hamiltonian.  Consequently,
diagonalization of $H^{(L=0)}$ gives the exact eigenvalues of the full
Hamiltonian belonging to the $L=0$ sector. 
The binary encoding maps
\begin{equation}
	|L=0,\alpha\rangle
	\longrightarrow
	|\mathrm{bin}(\alpha)\rangle ,
\end{equation}
so that the encoded Hamiltonian is represented in the corresponding
qubit Hilbert space.  Once represented as a matrix in the qubit basis,
it may, in principle, be decomposed into Pauli operators according to
\begin{equation}
	\hat H_{\rm q}
	=
	\sum_j c_jP_j,
	\qquad
	P_j\in\{I,X,Y,Z\}^{\otimes N_q}.
	\label{eq:H_qubit_pauli}
\end{equation}

The binary encoding itself introduces no approximation.  It changes
only the representation of the physical $L=0$ basis.  The physical
Hamiltonian is still the projected IBM Hamiltonian, with the
unphysical computational states excluded from the physical subspace.

\paragraph*{Symmetry-preserving variational space.}
\label{subsec:symmetry_aware_ansatz}
The symmetry reduction defines the physical variational space as the
$L=0$ sector. The production VQE subsequently parameterizes normalized
states within this space, as described in Sec.~\ref{sec:VQE}.
 Rather than starting
from a generic qubit ansatz that explores the entire computational
Hilbert space, the variational state is constructed entirely from the
physical $L=0$ basis,
\begin{equation}
	|\psi(\boldsymbol{\theta})\rangle
	=
	\sum_{\alpha=0}^{D_{L=0}-1}
	c_\alpha(\boldsymbol{\theta})
	|L=0,\alpha\rangle ,
	\label{eq:symmetry_vqe_ansatz}
\end{equation}
with $\sum_\alpha |c_\alpha(\boldsymbol{\theta})|^2 =1$. This construction guarantees
$ \langle\psi(\boldsymbol{\theta})|	\hat L^2 |\psi(\boldsymbol{\theta})\rangle = 0 $
for every value of the variational parameters.  The variational
optimization therefore cannot leave the physical $L=0$ sector.

The simplest examples illustrate the physical meaning of this
construction.  For $N_B=2$, the relevant $L=0$ configurations are
\begin{equation}
	|s^2\rangle,
	\qquad
	|d^2;L=0\rangle ,
\end{equation}
and the variational state can be written as
\begin{equation}
	|\psi(\theta)\rangle
	=
	\cos\theta\,|s^2\rangle
	+
	\sin\theta\,|d^2;L=0\rangle .
	\label{eq:NB2_ansatz}
\end{equation}
For $N_B=3$, the corresponding physically motivated $L=0$
configurations are
\begin{equation}
	|s^3\rangle,
	\qquad
	|s d^2;L=0\rangle,
	\qquad
	|d^3;L=0\rangle ,
\end{equation}
leading to a three-dimensional symmetry-reduced variational space.
The same construction is generalized to larger boson numbers by
retaining all independent $L=0$ configurations.  Importantly, these
configurations are not eigenstates of $\hat n_d$ in general, and the
VQE retains the required mixing between different $n_d$ sectors.
Thus, the symmetry-aware ansatz exploits the exact rotational symmetry
without imposing an additional, and generally incorrect, $U(5)$
$n_d$ conservation.

This construction is particularly advantageous compared with a generic
unconstrained qubit ansatz.  A generic ansatz can generate components
belonging to several angular-momentum sectors and must therefore learn
the symmetry through energy minimization.  In contrast, the present
symmetry-aware ansatz satisfies the angular-momentum constraint by
construction.

\paragraph*{Validation of the symmetry reduction.}
\label{subsec:symmetry_validation}

The symmetry reduction was validated by comparing the full IBM
Hamiltonian with its explicitly constructed $L=0$ projection.  For
each benchmark boson number, the lowest eigenvalue of the full IBM
Hamiltonian agrees with the lowest eigenvalue obtained in the
symmetry-reduced space, $ E_0^{\rm full} = 	E_0^{L=0} $ within numerical precision.

For $N_B=2,3,4,5,$ and $6$, the maximum absolute difference between the
full-space and $L=0$ ground-state energies was
\begin{equation}
	\max_{N_B}
	\left|
	E_0^{\rm full}-E_0^{L=0}
	\right|
	=
	1.22\times10^{-15}.
	\label{eq:L0_energy_validation}
\end{equation}

The overlap between the full-space ground state and its projection onto
the $L=0$ sector was unity to numerical precision, with the maximum
probability outside the $L=0$ sector equal to only
$
	P_{\rm outside}^{\rm max}
	=
	3.33\times10^{-16}.
	\label{eq:L0_probability_validation}
$

The symmetry-reduced VQE was then tested against the same exact
ground states.  For all values $N_B=2$--$6$, the minimum VQE fidelity
was $ F_{\rm min}=1.000000000000000 $, while the maximum energy error was
$ \max|\Delta E_{\rm VQE}| =1.83\times10^{-15} $.
The maximum deviation of the $d$-boson number expectation value was
$ \max|\Delta\langle n_d\rangle| = 5.24\times10^{-9}$,
and the corresponding $L^2$ expectation value remained zero within
numerical precision.

These results demonstrate that the symmetry-aware variational
construction reproduces the exact IBM ground state while operating in
a drastically smaller Hilbert space.

\paragraph*{Resource reduction and validation chain.}
\label{subsec:resource_validation_chain}

The complete computational procedure can therefore be summarized as
\begin{equation}
			\mathcal H_{\rm IBM}
		\;\longrightarrow\;
		\mathcal H_{L=0}
		\;\longrightarrow\;
		\mathcal H_{\rm enc}
		\;\longrightarrow\;
		|\psi(\boldsymbol{\theta})\rangle
		\;\longrightarrow\;
		E_{\rm VQE}.
		\label{eq:validation_chain}
\end{equation}

The first step removes only states belonging to symmetry sectors that
cannot contribute to the $L=0$ ground state.  The second step provides
an exact binary representation of the remaining physical basis.  The
third step constructs a variational state entirely within this
symmetry-preserving space.

The numerical validation shows that this sequence introduces no
observable loss of physical information for the ground-state problem
studied here.  The symmetry reduction therefore simultaneously reduces
the quantum-resource requirements and provides a physically informed
variational manifold for the VQE calculation.

\section{Variational quantum eigensolver}
\label{sec:VQE}

After constructing the symmetry-reduced IBM Hamiltonian and its
minimum-qubit encoding, the ground-state problem is solved using the
variational quantum eigensolver (VQE). The VQE provides a variational
approximation to the lowest eigenvalue of the encoded Hamiltonian, while
the optimized state allows the ground-state observables relevant to the
quantum phase transition to be evaluated.

The VQE calculation is performed directly in the symmetry-reduced
$L=0$ sector.  The reduced Hamiltonian $H_{L=0}$ is therefore mapped
to a minimum-qubit register according to
\begin{equation}
	H_{L=0}\longrightarrow H_q ,
	\label{eq:reduced_to_qubit}
\end{equation}
where $H_q$ denotes the qubit representation of the physical
symmetry-reduced Hamiltonian.  The variational optimization is then
carried out over states represented in this reduced physical space,
rather than over the complete IBM Hilbert space.  Thus, the quantum
resource requirements are determined by $D_{L=0}$ rather than by the
full IBM dimension $D_{\rm IBM}$.

No hardware execution is performed in the present study; the VQE is
used as a numerically controlled variational benchmark of the
symmetry-reduced IBM problem.

\paragraph*{Variational principle and trial state.}
\label{subsec:vqe_variational_principle}
For a trial state $|\psi(\boldsymbol{\theta})\rangle$ in the encoded
reduced space, the VQE minimizes the expectation value
\begin{equation}
	E(\boldsymbol{\theta})
	=
	\frac{
		\langle\psi(\boldsymbol{\theta})|
		H_q
		|\psi(\boldsymbol{\theta})\rangle
	}{
		\langle\psi(\boldsymbol{\theta})
		|\psi(\boldsymbol{\theta})\rangle
	},
	\label{eq:vqe_energy}
\end{equation}
with respect to the variational parameters $\boldsymbol{\theta}$.
For normalized trial states, this reduces to
\begin{equation}
	E(\boldsymbol{\theta})
	=
	\langle\psi(\boldsymbol{\theta})|
	H_q
	|\psi(\boldsymbol{\theta})\rangle .
\end{equation}
The variational ground-state energy is consequently obtained as
\begin{equation}
	E_{\rm VQE}
	=
	\min_{\boldsymbol{\theta}}
	E(\boldsymbol{\theta}) .
	\label{eq:vqe_minimization}
\end{equation}

For the QPT calculations, the VQE is performed directly in the
symmetry-reduced $L=0$ Hilbert space. If its dimension is $D_{L=0}$,
the normalized variational state is parameterized directly by
$D_{L=0}-1$ independent angular variables. Consequently, the
variational optimization is carried out only within the physical
symmetry sector, without introducing variational degrees of freedom
associated with other angular-momentum sectors.

\paragraph*{Direct reduced-space state parameterization.}
\label{subsec:vqe_state_parameterization}

Unlike a fixed gate-level ansatz, the production VQE calculations use
a direct parameterization of the normalized state vector in the
$D_{L=0}$-dimensional reduced Hilbert space. The state is constructed
from $D_{L=0}-1$ angular parameters,
\begin{equation}
	\boldsymbol{\theta}
	=
	(\theta_1,\ldots,\theta_{D_{L=0}-1}),
\end{equation}
with the corresponding normalized amplitudes generated by a standard
hyperspherical-coordinate parameterization. This provides a complete
parameterization of a real normalized state in the reduced space, up
to an irrelevant global phase.

The use of the reduced space is essential for the present application:
the number of variational parameters scales with $D_{L=0}$ rather than
with the dimension of the complete IBM Hilbert space. For the boson
numbers considered here,
\[
D_{L=0} \ll D_{\rm IBM},
\]
so that the symmetry reduction substantially decreases the variational
search space.

This direct reduced-space parameterization is used for the production
QPT calculations reported below. Gate-level ansatz constructions are
not required for these calculations because the objective is to
benchmark the symmetry-reduced variational formulation and its ability
to reproduce the finite-size IBM quantum phase transition.

\paragraph*{Hamiltonian representation and Pauli decomposition.}
\label{subsec:vqe_energy_evaluation}

The symmetry-reduced Hamiltonian is represented directly as the
$D_{L=0}\times D_{L=0}$ matrix $H_{L=0}$ for the production VQE
calculations. The variational energy is therefore evaluated as
\begin{equation}
	E_{\rm VQE}(\boldsymbol{\theta})
	=
	\langle\psi(\boldsymbol{\theta})|
	H_{L=0}
	|\psi(\boldsymbol{\theta})\rangle .
	\label{eq:vqe_reduced_energy}
\end{equation}

For completeness, the corresponding minimum-qubit representation can
be expressed as a linear combination of Pauli strings,
\begin{equation}
	H_q=\sum_j c_j P_j,
	\qquad
	P_j\in\{I,X,Y,Z\}^{\otimes N_q},
\end{equation}
which establishes the connection between the reduced IBM problem and
a quantum-computing implementation. The Pauli decomposition is used as
an independent check of the encoded Hamiltonian and is not required for
the direct reduced-space VQE optimization reported here.
\paragraph*{Optimization strategy.}
\label{subsec:vqe_optimization}

The variational parameters are determined by numerical minimization of
Eq.~\eqref{eq:vqe_energy}. A gradient-based BFGS minimization is
performed for each initial parameter configuration.

To reduce sensitivity to the initial state and possible local minima,
the optimization is repeated from multiple starting configurations.
The zero-parameter configuration and a set of independent random
initializations are included. In addition, when the exact reduced
ground state is available for validation, its hyperspherical
coordinates are supplied as an additional starting configuration.
For each calculation, the converged solution having the lowest
variational energy is retained.

For the production QPT calculations, the exact ground state is not used
to initialize the variational optimization. Twenty independent random
initializations and the zero-parameter configuration are employed, and
the converged solution with the lowest variational energy is retained.
The exact reduced ground state is used only in separate validation
calculations.

\paragraph*{Ground-state validation.}
\label{subsec:vqe_ground_state_validation}

The VQE results are validated against exact diagonalization of the same
symmetry-reduced Hamiltonian. The energy difference is defined by
\begin{equation}
	\Delta E
	=
	E_{\rm VQE}
	-
	E_{\rm exact}.
	\label{eq:vqe_energy_error}
\end{equation}
For a properly converged variational calculation,
$\Delta E\geq0$ apart from numerical round-off and optimization
tolerances.

The agreement between the VQE and exact ground states is quantified by
the fidelity
\begin{equation}
	F
	=
	\left|
	\langle\psi_{\rm exact}
	|\psi_{\rm VQE}\rangle
	\right|^2 .
	\label{eq:vqe_fidelity}
\end{equation}
A value of $F=1$ corresponds to identical states up to an irrelevant
global phase.

The physical character of the optimized state is additionally checked
through the probability contained in the physical encoded subspace,
\begin{equation}
	P_{\rm phys}
	=
	\langle\psi_{\rm VQE}|
	\hat P_{\rm phys}
	|\psi_{\rm VQE}\rangle .
	\label{eq:vqe_physical_probability}
\end{equation}

For the direct reduced-space VQE calculations reported in the QPT
analysis, $P_{\rm phys}=1$ identically by construction. For the
embedded minimum-qubit representation, the corresponding physical
probability provides an additional check that the penalty construction
successfully separates the unused computational states.

\paragraph*{VQE ground-state observables.}
\label{subsec:vqe_observables}

Once the variational ground state has been obtained, the same
observables used in the exact calculation are evaluated from the VQE
state. In particular, the normalized $d$-boson occupation is
\begin{equation}
	\rho_d^{\rm VQE}
	=
	\frac{
		\langle\psi_{\rm VQE}|
		\hat n_d
		|\psi_{\rm VQE}\rangle
	}{N_B},
	\label{eq:vqe_rho_d}
\end{equation}
and the normalized quadrupole correlation is
\begin{equation}
	C_Q^{\rm VQE}
	=
	\frac{
		\langle\psi_{\rm VQE}|
		\hat Q_\chi\cdot\hat Q_\chi
		|\psi_{\rm VQE}\rangle
	}{N_B^2}.
	\label{eq:vqe_Q_correlation}
\end{equation}

The rotational character of the optimized state is further verified
through
\begin{equation}
	\langle L^2\rangle_{\rm VQE}
	=
	\langle\psi_{\rm VQE}|
	\hat L^2
	|\psi_{\rm VQE}\rangle .
	\label{eq:vqe_L2}
\end{equation}
Because the VQE is performed directly in the $L=0$ sector, this
quantity vanishes up to numerical precision.

The agreement of $\rho_d^{\rm VQE}$ and $C_Q^{\rm VQE}$ with their exact
counterparts is particularly important here, since the objective is
not only to reproduce the ground-state energy but also the structural
information contained in the ground-state wave function. These
observables are subsequently used to identify the finite-size
signatures of the $U(5)\rightarrow SU(3)$ quantum phase transition.

\paragraph*{VQE reproduction of the finite-size quantum phase transition.}
\label{subsec:vqe_qpt_reproduction}

Having established the accuracy of the variational ground states, the
VQE is applied throughout the control-parameter range $ 0\leq\xi\leq1 $
at fixed
$\chi=-\sqrt{7}/2$.
The VQE ground-state energy and structural observables are evaluated
on the same $\xi$ grid used for exact diagonalization. The resulting
finite-$N_B$ indicators are then compared directly with the exact
results.

Two complementary finite-size analyses are performed. The first
tracks the maximum slope of $\rho_d$ and extrapolates its
pseudocritical position. The second independently defines
energy- and order-parameter-based pseudocritical indicators and
analyzes their extrapolations.

The resulting agreement demonstrates that the reduced-space VQE
reproduces not only the ground-state energy but also the structural
information required to characterize the
$U(5)\rightarrow SU(3)$ quantum phase transition.
\section{VQE description of the quantum phase transition}
\label{sec:VQE_QPT}
	
\begin{figure}[p]
		\centering
		
		\subfloat[Ground-state energy.]
		{
			\includegraphics[width=0.47\textwidth]{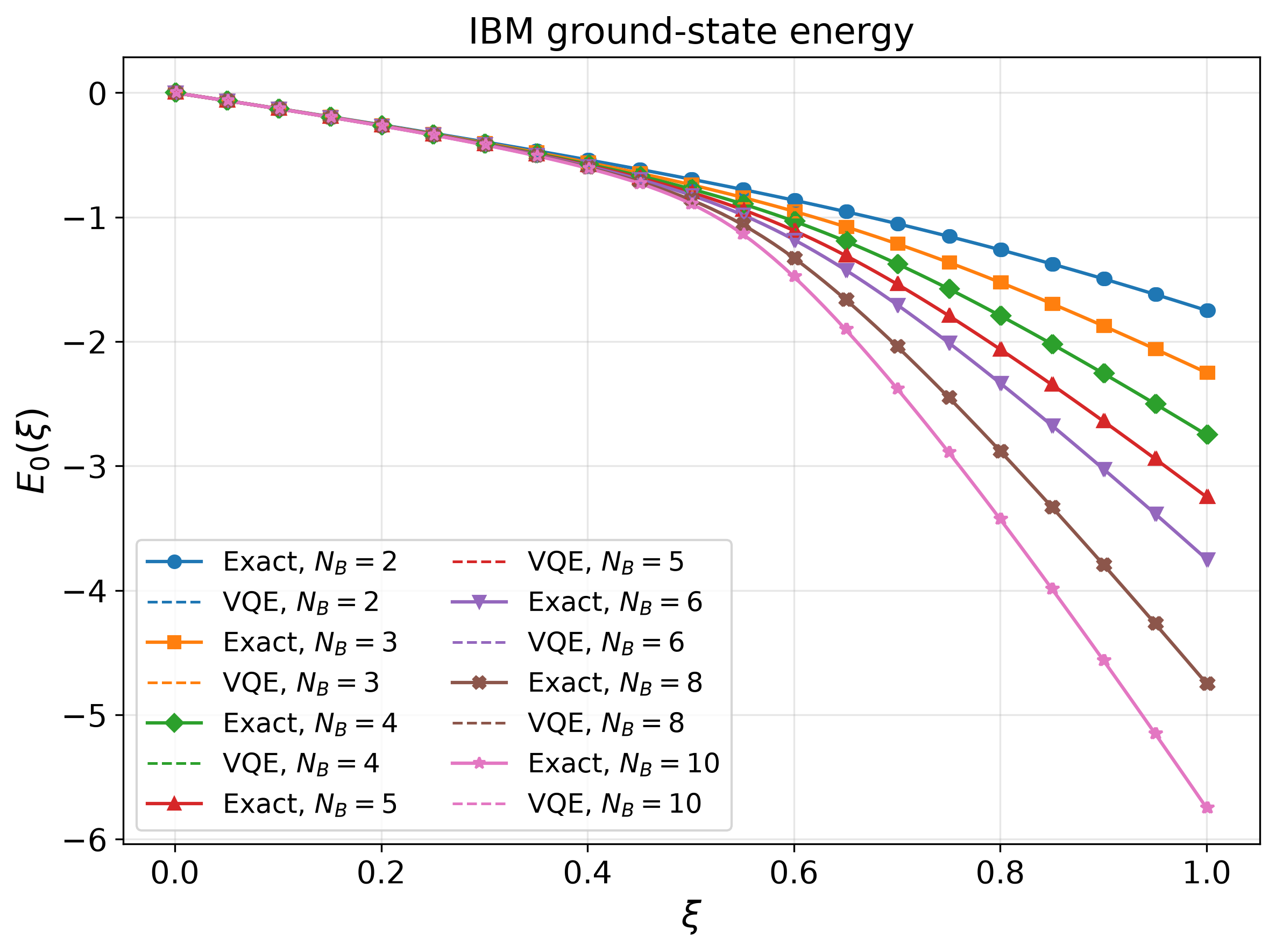}
			\label{fig:qpt_energy}
		}
		\hfill
		\subfloat[Normalized $d$-boson occupation.]
		{
			\includegraphics[width=0.47\textwidth]{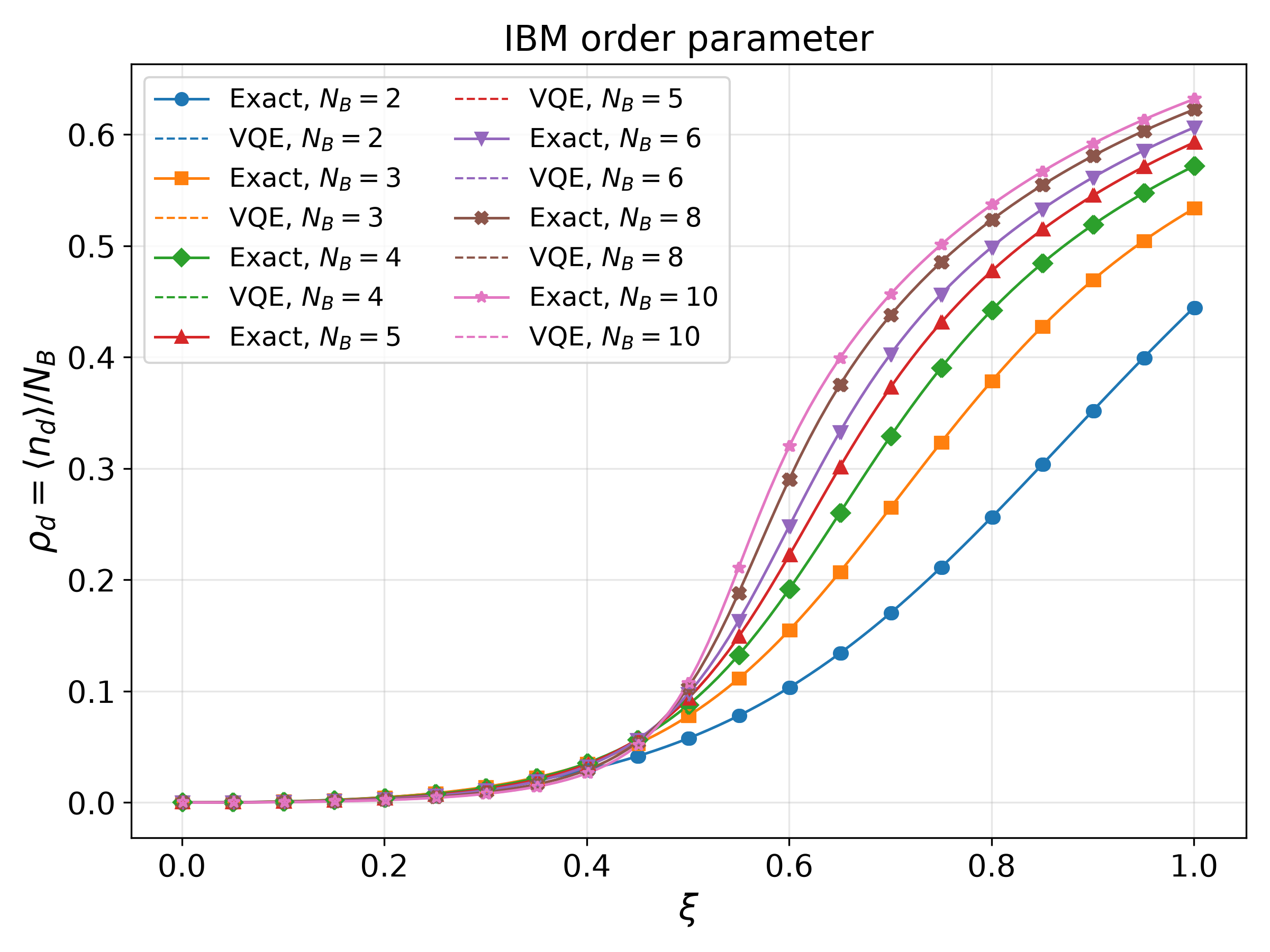}
			\label{fig:qpt_order_parameter}
		}
		
		\vspace{0.5cm}
		
		\subfloat[First derivative of the ground-state energy.]
		{
			\includegraphics[width=0.47\textwidth]{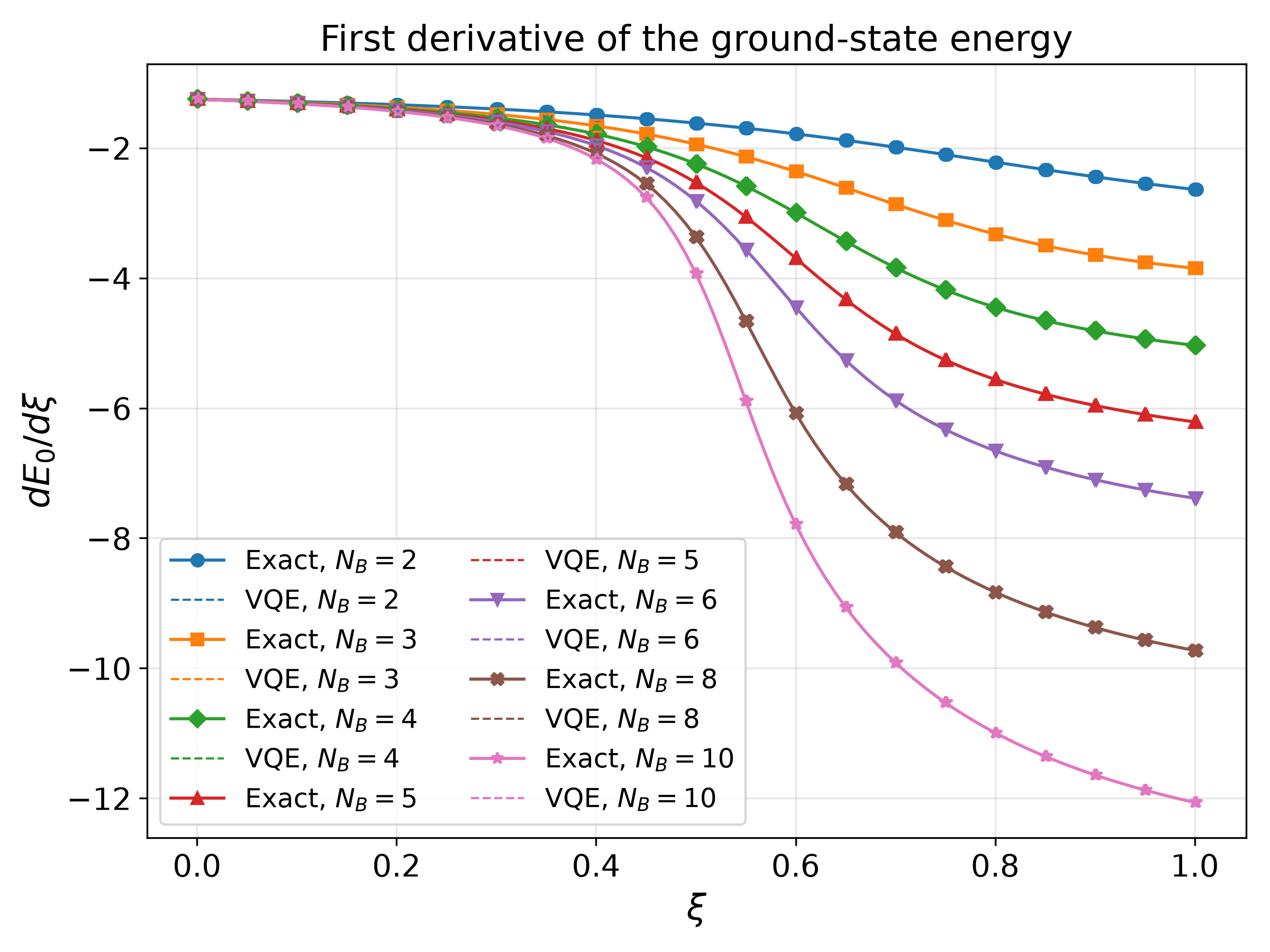}
			\label{fig:qpt_first_derivative}
		}
		\hfill
		\subfloat[Second derivative of the ground-state energy.]
		{
			\includegraphics[width=0.47\textwidth]{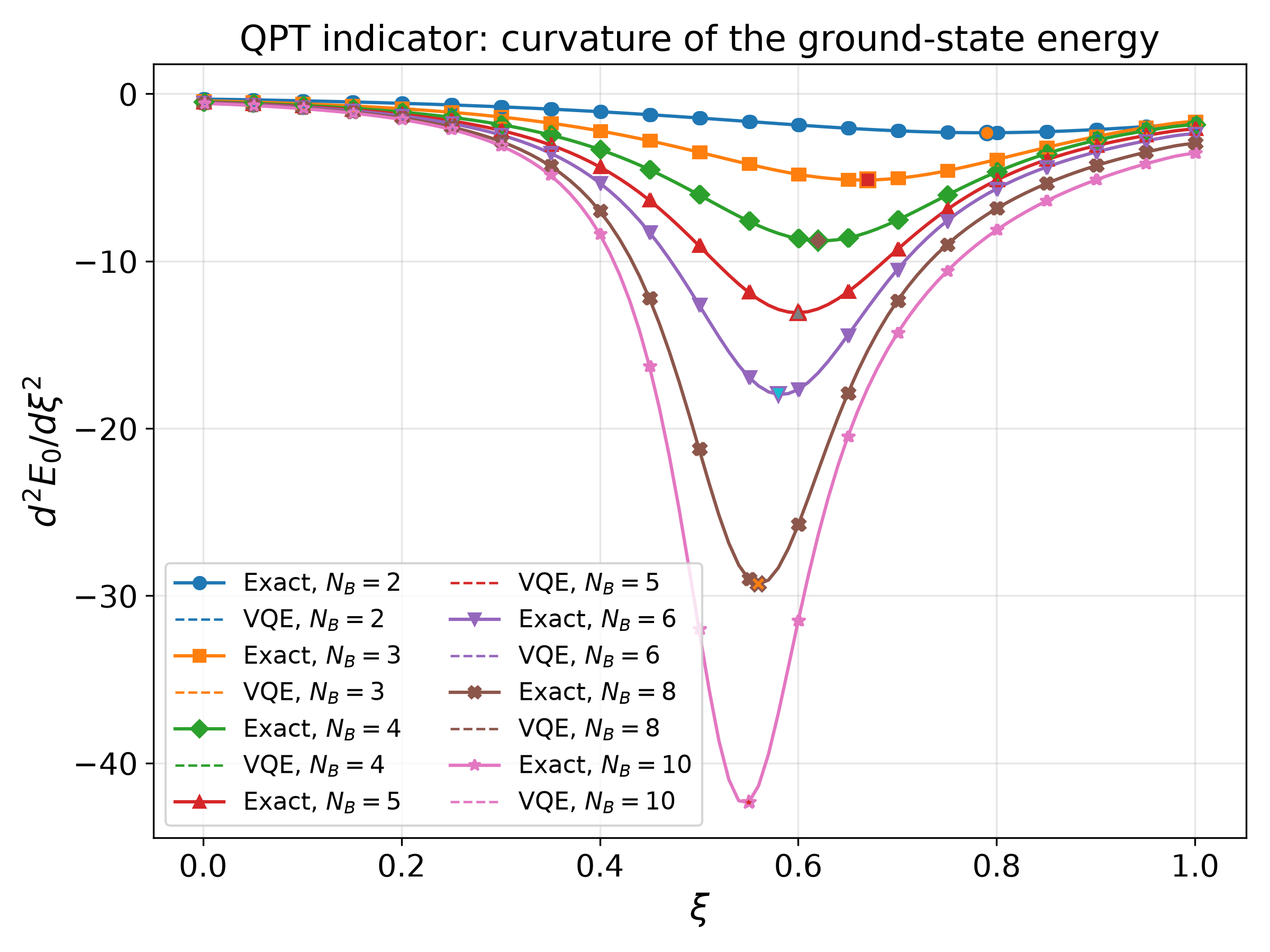}
			\label{fig:qpt_second_derivative}
		}
		
		\caption{
			Finite-size evolution of the $U(5)\rightarrow SU(3)$ quantum
			phase transition in the IBM. Panels (a) and (b) show the
			ground-state energy and normalized $d$-boson occupation,
			respectively, while panels (c) and (d) show the first and
			second derivatives of the ground-state energy. Results are
			shown for the investigated values of $N_B$. The derivative
			responses provide complementary indicators of the
			finite-size pseudocritical region.
		}
		\label{fig:qpt_four_panel}
	\end{figure}

Following validation against exact diagonalization, the ability of the VQE calculation to reproduce the finite-size quantum phase transition is next examined.
 Before introducing the two independent indicators, 
	the finite-size evolution of the transition is illustrated in
	Fig.~\ref{fig:qpt_four_panel}. The ground-state energy and normalized
	$d$-boson occupation exhibit the expected smooth finite-size evolution
	across the transition, while their derivative responses become
	increasingly pronounced with increasing $N_B$. These complementary
	responses provide the basis for defining independent
	energy- and order-parameter-based pseudocritical indicators.
	
	Two independent indicators are considered:
	the energy-based finite-size pseudocritical indicator, $\xi_E(N_B)$, and the order-parameter-based finite-size pseudocritical indicator, $\xi_\rho(N_B)$. 
		
	For each finite boson number $N_B$, two independent pseudocritical
	indicators are extracted from the energy and order-parameter responses.
	The energy-based indicator, $\xi_E(N_B)$, is defined as the value of
	$\xi$ at which the curvature of the ground-state energy is maximal,
	\begin{equation}
		\xi_E(N_B)
		=
		\operatorname*{arg\,max}_{\xi}
		\left[
		-\frac{\partial^2 E_0(\xi,N_B)}
		{\partial \xi^2}
		\right].
		\label{eq:xi_E_definition}
	\end{equation}
	The order-parameter indicator, $\xi_\rho(N_B)$, is defined from the
	maximum response of the normalized $d$-boson occupation,
	\begin{equation}
		\rho(\xi,N_B)
		=
		\frac{\langle n_d\rangle}{N_B},
	\end{equation}
	as
	\begin{equation}
		\xi_\rho(N_B)
		=
		\operatorname*{arg\,max}_{\xi}
		\left[
		\frac{\partial \rho(\xi,N_B)}
		{\partial \xi}
		\right].
		\label{eq:xi_rho_definition}
	\end{equation}
	Thus, $\xi_E$ and $\xi_\rho$ provide independent finite-size
	estimates of the transition region based, respectively, on the
	ground-state energy curvature and the order-parameter response.

\begin{figure}[t]
		\centering
		\includegraphics[width=0.80\textwidth]{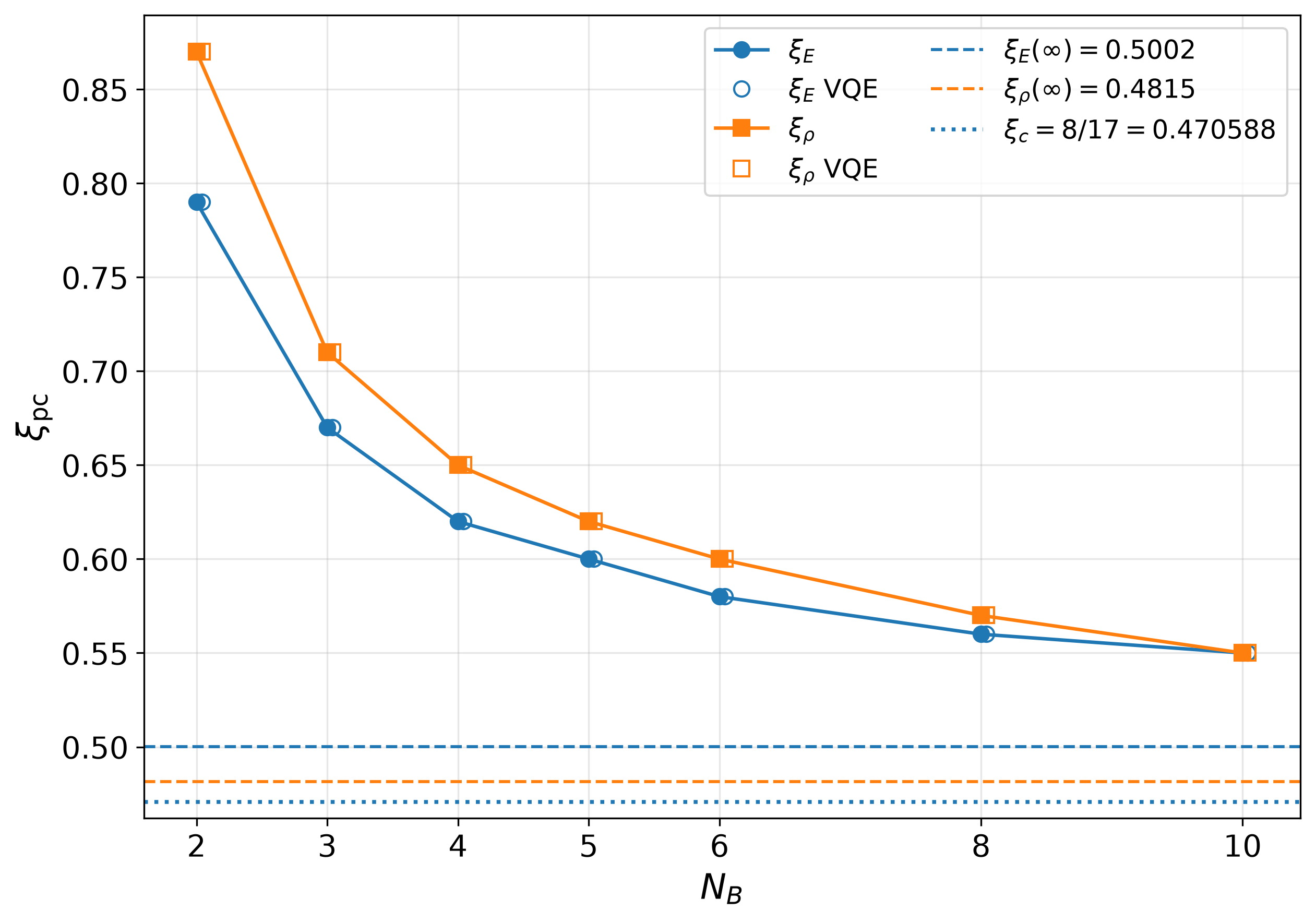}
		\caption{
			Finite-$N_B$ pseudocritical indicators $\xi_E$ and $\xi_\rho$
			obtained from the energy and order-parameter responses,
			respectively. The extrapolated thermodynamic-limit values are
			compared with the analytic IBM critical point
			$\xi_c=8/17\simeq0.470588$. Exact and VQE results are
			indistinguishable on the scale of the figure.
		}
		\label{fig:qpt_final_critical_points}
	\end{figure}

	The finite-$N_B$ results are summarized in
	Fig.~\ref{fig:qpt_final_critical_points}.  For small boson numbers the
	two criteria exhibit a visible separation.  This difference decreases
	systematically with increasing $N_B$, reaching the same value at
	$N_B=10$ for the present numerical grid.	
	
	\paragraph*{Finite-$N_B$ QPT indicators}
	
	The exact and VQE values are identical for all benchmark boson numbers.
	The results are listed in Table~\ref{tab:final_qpt}. 
		The separation between the two indicators decreases systematically
	with increasing boson number. This behavior is consistent with the
	interpretation that different finite-size definitions of the
	transition converge toward a common thermodynamic-limit region.	
	\begin{table}[ht]
		\centering
		\caption{Finite-$N_B$ QPT indicators extracted from the exact and VQE
			calculations.}
		\label{tab:final_qpt}
		\begin{tabular}{cccccc}
			\toprule
			$N_B$ & $\xi_E$ & $\xi_\rho$ & $|\Delta\xi|$
			& $\xi_E^{\rm VQE}$ & $\xi_\rho^{\rm VQE}$\\
			\midrule
			2  & 0.79 & 0.87 & 0.08 & 0.79 & 0.87\\
			3  & 0.67 & 0.71 & 0.04 & 0.67 & 0.71\\
			4  & 0.62 & 0.65 & 0.03 & 0.62 & 0.65\\
			5  & 0.60 & 0.62 & 0.02 & 0.60 & 0.62\\
			6  & 0.58 & 0.60 & 0.02 & 0.58 & 0.60\\
			8  & 0.56 & 0.57 & 0.01 & 0.56 & 0.57\\
			10 & 0.55 & 0.55 & 0.00 & 0.55 & 0.55\\
			\bottomrule
		\end{tabular}
	\end{table}

\begin{figure}[t]
		\centering
		\includegraphics[width=0.72\textwidth]{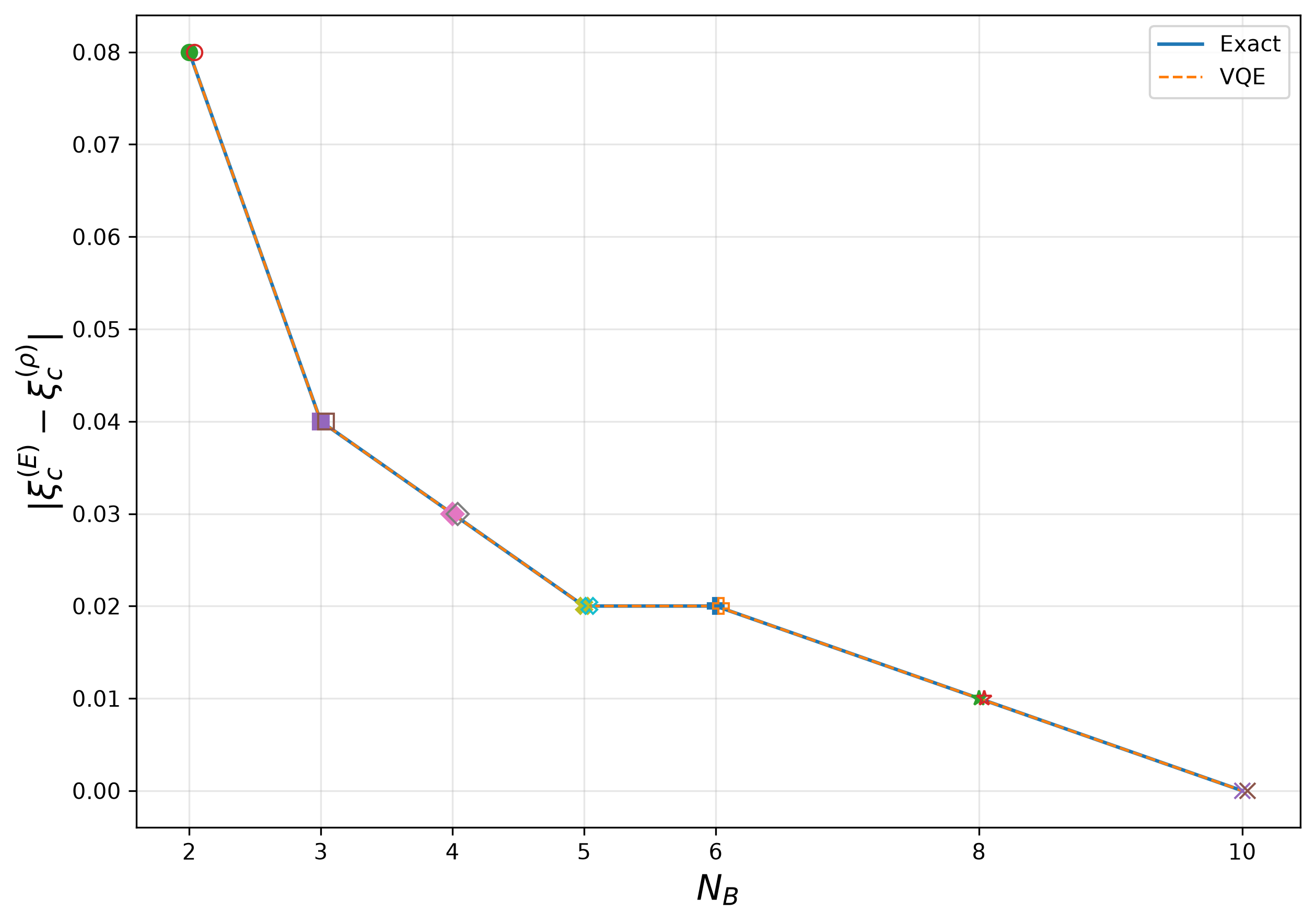}
		\caption{
			Difference between the energy- and order-parameter-based
			finite-$N_B$ QPT indicators,
			$\Delta\xi=|\xi_E-\xi_\rho|$.
			The decreasing difference demonstrates the increasing consistency
			of the two criteria with increasing boson number.
		}
		\label{fig:qpt_final_indicator_difference}
	\end{figure}

	The difference between the two finite-size indicators provides a useful
	measure of the ambiguity associated with identifying a transition in a
	finite system.  
	The quantity
	\begin{equation}
		\Delta\xi(N_B)
		=
		|\xi_E(N_B)-\xi_\rho(N_B)|
		\label{eq:indicator_difference}
	\end{equation}
	is therefore defined.
	The resulting behavior is displayed in
	Fig.~\ref{fig:qpt_final_indicator_difference}.  The indicator
	difference decreases from $0.08$ at $N_B=2$ to zero at $N_B=10$.
	Thus, the energy and order-parameter criteria become progressively more
	consistent as the boson number increases.

\begin{figure}[t]
	\centering
	\includegraphics[width=0.80\textwidth]{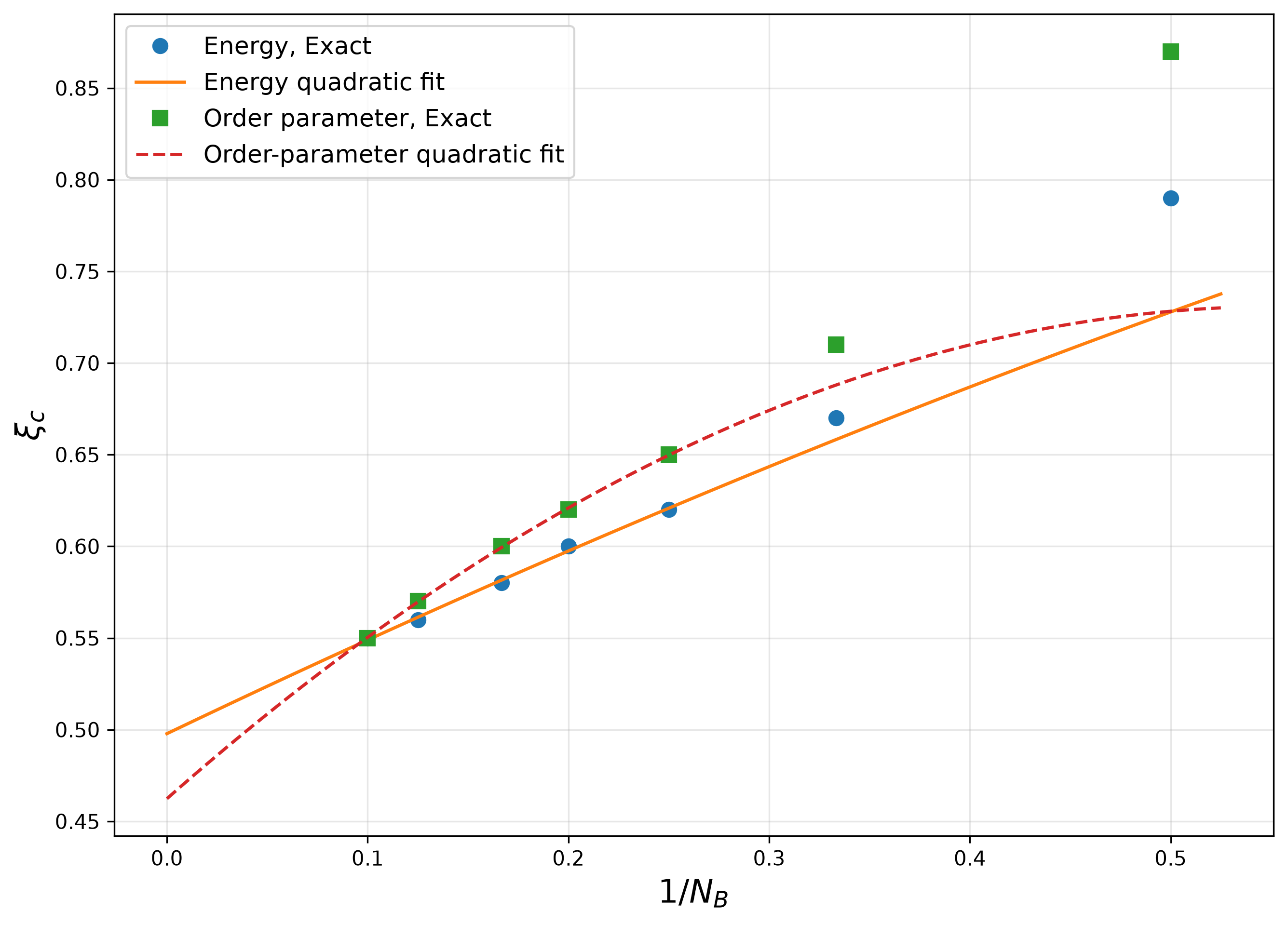}
	\caption{
		Finite-size scaling and thermodynamic-limit extrapolations of the
		QPT indicators.  Linear and quadratic fits in inverse boson number
		are shown for the energy and order-parameter criteria.  The spread
		among the extrapolations is used to quantify the systematic
		uncertainty associated with the finite-size extrapolation.
	}
	\label{fig:qpt_final_finite_size_scaling}
\end{figure}

The finite-size evolution of the critical indicators is further examined
through thermodynamic-limit extrapolations.  Both linear and quadratic
forms in inverse boson number are considered.  The resulting
extrapolations are shown in Fig.~\ref{fig:qpt_final_finite_size_scaling}.
For the $\xi_E(N_B)$, the extrapolated values obtained from the six
fits span $ 0.48155 \lesssim \xi_E(\infty) \lesssim 0.51517 $,
whereas the $\xi_\rho(N_B)$ gives
$	0.46233	\lesssim\xi_\rho(\infty) \lesssim 0.50595 $.

The spread of the extrapolations provides a conservative estimate of the
systematic uncertainty associated with the finite-size fitting
procedure.

\paragraph*{Exact--VQE agreement.}
	
	For the finite-size QPT indicators, the quantities $ \max|\xi_E^{\rm VQE}-\xi_E^{\rm exact}| $ and
	$ \max|\xi_\rho^{\rm VQE}-\xi_\rho^{\rm exact}|$
	 are found to be identical to the numerical precision of the stored analysis data.
		Thus, the VQE reproduces the complete set of extracted finite-size
	critical indicators.
\section{Thermodynamic-limit analysis}
\label{sec:thermodynamic_limit}
	
	The finite-$N_B$ indicators do not converge monotonically to exactly
	the same value over the relatively small boson-number range accessible
	to the present calculation. An explicit thermodynamic-limit extrapolation is therefore performed.
		
	For each indicator, both linear and quadratic forms in the inverse boson number are considered,	
	\begin{equation}
		\xi(N_B)
		=
		\xi_\infty+\frac{a}{N_B},
		\label{eq:linear_extrapolation}
	\end{equation}
	and
	\begin{equation}
		\xi(N_B)
		=
		\xi_\infty+\frac{a}{N_B}+\frac{b}{N_B^2}.
		\label{eq:quadratic_extrapolation}
	\end{equation}
	
	The extrapolations were performed over the ranges
	$2$--$10$, $3$--$10$, and $4$--$10$.	
	The resulting thermodynamic-limit extrapolations are summarized in
	Table~\ref{tab:thermodynamic_extrapolation}.
	
	\begin{table*}[ht]
		\centering
		\caption{Thermodynamic-limit extrapolations of the energy and
			order-parameter QPT indicators.}
		\label{tab:thermodynamic_extrapolation}
		\begin{tabular}{ccccc}
			\toprule
			Range & Fit & Indicator & $\xi_\infty$ & $R^2$\\
			\midrule
			2--10 & linear    & $E$   & 0.4815477 & 0.989945\\
			2--10 & quadratic & $E$   & 0.5151677 & 0.999408\\
			2--10 & linear    & $\rho$ & 0.4657265 & 0.990655\\
			2--10 & quadratic & $\rho$ & 0.5059462 & 0.998540\\
			3--10 & linear    & $E$   & 0.4962998 & 0.996063\\
			3--10 & quadratic & $E$   & 0.5091615 & 0.998050\\
			3--10 & linear    & $\rho$ & 0.4849075 & 0.998497\\
			3--10 & quadratic & $\rho$ & 0.4835866 & 0.998509\\
			4--10 & linear    & $E$   & 0.5013184 & 0.996100\\
			4--10 & quadratic & $E$   & 0.4978033 & 0.996239\\
			4--10 & linear    & $\rho$ & 0.4863477 & 0.996330\\
			4--10 & quadratic & $\rho$ & 0.4623273 & 0.999736\\
			\bottomrule
		\end{tabular}
	\end{table*}
	
    The extrapolated energy indicator yields a mean value of $\overline{\xi}_E = 0.5002164$ and a standard deviation of $\sigma_E = 0.0106023$.
	The corresponding order-parameter indicator has $ \overline{\xi_\rho} =	0.4814736 $ with $ 	\sigma_\rho=0.0144538$.
	The half-ranges of the extrapolated values are $\delta_E=0.0168100$	and $\delta_\rho=0.0218095$.	
	The mean of the extrapolated values is therefore used as the central estimate of the final thermodynamic-limit indicators, while half of their total spread is taken as a measure of the systematic uncertainty associated with the fitting range
	\begin{equation}
			\xi_E(\infty)
			=
			0.5002\pm0.0168,
		\label{eq:xiE_final}
	\end{equation}
	and
	\begin{equation}
			\xi_\rho(\infty)
			=
			0.4815\pm0.0218.
		\label{eq:xirho_final}
	\end{equation}
	
\begin{figure}[t]
	\centering
	\includegraphics[width=0.72\textwidth]{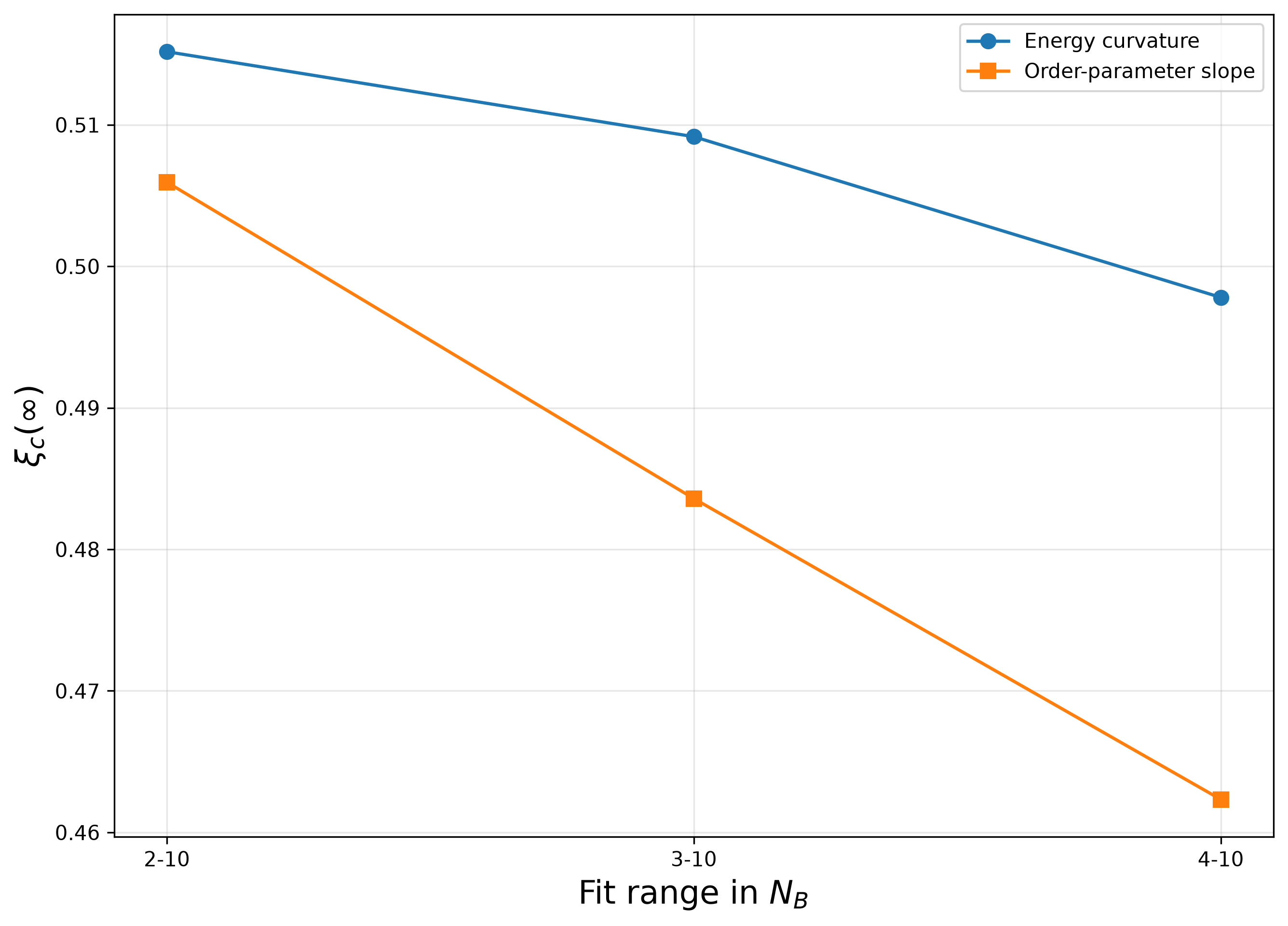}
	\caption{
		Thermodynamic-limit estimates of the critical control parameter
		obtained from the energy and order-parameter criteria.  The central
		values are obtained by averaging over the investigated finite-size
		extrapolations, while the quoted uncertainties represent the
		corresponding half-range estimates.
	}
	\label{fig:qpt_final_thermodynamic_limit}
\end{figure}

	The stability of the thermodynamic-limit estimates is summarized in
	Fig.~\ref{fig:qpt_final_thermodynamic_limit}.  Averaging over the
	investigated fitting ranges gives $	\xi_E(\infty) = 0.50022\pm0.01681$ and $ \xi_\rho(\infty) = 0.48147\pm0.02181$.	
The difference between these two independent estimates is
$ 	|\Delta\xi_\infty|	=	0.01874	$.
The two estimates therefore overlap within the systematic uncertainty
associated with the finite-size extrapolation.  This agreement provides
evidence that the two independent observables identify the same
transition region in the thermodynamic limit.

	\paragraph*{Stability of the extrapolation.}
	
	The spread of the extrapolated values provides a useful measure of the
	sensitivity to the finite-size fitting procedure. For $\xi_E(N_B)$ the extrapolations span
	$ 0.48155\leq\xi_E(\infty)\leq0.51517 $
	whereas $\xi_\rho(N_B)$ spans
	$ 0.46233\leq\xi_\rho(\infty)\leq0.50595 $.
	The larger spread of the order-parameter extrapolations emphasizes the
	importance of treating the thermodynamic-limit value as an estimate
	rather than as a single exact number when only the finite boson
	numbers $N_B=2$--$10$ are available.

	\paragraph*{Relation to the analytic critical point.}
	
	The analytic coherent-state result,
	$ \xi_c={8}/{17}=0.470588 $,
	provides the exact thermodynamic-limit benchmark for the IBM
	Hamiltonian considered here.

	The energy-based extrapolation,
	$\xi_E(\infty)=0.5002\pm0.0168$, lies above the analytic critical
	point, although it remains close to the finite-size critical region.
	The order-parameter extrapolation,
	$\xi_\rho(\infty)=0.4815\pm0.0218$, is consistent with the analytic
	value within the quoted finite-size spread.
	The difference between the two indicators,
	$ |\Delta\xi_\infty|=0.01874 $,
	is small compared with the overall finite-size displacement observed
	at the smallest boson numbers.

	\paragraph*{Final QPT and VQE consistency analysis}
	\label{sec:final_validation}
	
	The final analysis combines the exact and variational calculations,
	finite-size QPT indicators, and thermodynamic-limit extrapolations.	
	For the seven boson numbers $ N_B=2,3,4,5,6,8,10$,
	the exact and VQE indicators, i.e.,$\max|\Delta\xi_E|$ and $\max|\Delta\xi_\rho|$ are identical within numerical precision of the stored data.

	The finite-size difference between the energy and order-parameter
	criteria decreases from $0.08 $ at $N_B=2$ to zero at $N_B=10$ on the present $\xi$ grid.
		This behavior provides a useful internal consistency check. The two
	criteria probe different properties of the ground state and need not
	coincide within numerical precision at finite boson number. Their convergence therefore
	provides evidence that the observed transition is not an artifact of a
	single diagnostic.

	A direct comparison of the exact and VQE results gives no visible
	difference on the scale of the QPT figures.	
	The final consistency analysis therefore establishes three distinct
	levels of agreement:
	
	\begin{enumerate}
		\item the encoded Hamiltonian reproduces the symmetry-reduced IBM
		Hamiltonian;
		\item VQE reproduces the exact ground-state energy and observables;
		\item VQE reproduces the finite-size QPT indicators and their
		thermodynamic-limit behavior.
	\end{enumerate}
	
	These tests demonstrate that the quantum algorithm is capable of
	capturing the structural information associated with the IBM quantum
	phase transition rather than merely minimizing an energy functional.

	The most direct test of the quantum simulation is provided by comparing
	the thermodynamic-limit extrapolations obtained independently from the
	exact and VQE data.  As shown in Fig.~\ref{fig:qpt_final_critical_points},
	the two calculations produce identical extrapolated values for both
	indicators. 	
	The maximum deviations $ \max |\xi_E^{\rm VQE}-\xi_E^{\rm exact}| $ and 
	$ \max 	|\xi_\rho^{\rm VQE}-\xi_\rho^{\rm exact}| $
	are almost identical within the numerical precision of the present analysis.
	
	This result is particularly significant because the agreement is not
	restricted to a single ground-state energy.  The complete finite-size
	QPT analysis, including the extraction of critical indicators and their
	thermodynamic-limit extrapolation, is reproduced by the VQE calculation.

	\paragraph*{Overall interpretation of the numerical results.}
	\label{subsec:overall_qpt_interpretation}
	
	The sequence of results presented above establishes the QPT from several
	complementary perspectives.  The evolution of $\rho_d$ in
	Fig.~\ref{fig:rho_d_NB10} demonstrates the microscopic structural
	change from the spherical to the deformed regime.  The finite-size
	scaling of the pseudocritical point in
	Fig.~\ref{fig:finite_size_xipc} shows that the crossover position
	converges systematically with increasing boson number, while the
	growth of $D_{\max}$ in Fig.~\ref{fig:slope_scaling} demonstrates the
	progressive sharpening of the transition.
	
	The final QPT analysis provides an independent and more conservative
	determination of the transition region.  As shown in
	Fig.~\ref{fig:qpt_final_critical_points}, the energy and order-parameter
	criteria approach one another with increasing $N_B$.  Their difference,
	shown in Fig.~\ref{fig:qpt_final_indicator_difference}, decreases
	systematically, indicating that the ambiguity associated with
	finite-size identification of the transition becomes smaller for
	larger systems.
	
	The thermodynamic-limit extrapolations in
	Fig.~\ref{fig:qpt_final_thermodynamic_limit} yield
	$ \xi_E(\infty)	= 0.50022\pm0.01681 $, and $ \xi_\rho(\infty) =	0.48147\pm0.02181$.
	Their difference, $	|\Delta\xi_\infty|=0.01874$,
	is smaller than the combined finite-size extrapolation uncertainty.
	The two criteria therefore identify a consistent transition region.
	
	Most importantly, Fig.~\ref{fig:qpt_final_critical_points} demonstrates
	that the same conclusions are obtained from the VQE calculation.
	Within the numerical precision of the present study, the VQE and exact
	results give identical finite-$N_B$ indicators and identical
	thermodynamic-limit extrapolations.  The quantum calculation therefore
	reproduces the complete structural analysis rather than merely the
	ground-state energy.
	
	\section{Discussion}
	\label{sec:discussion}
	
	The results demonstrate that the IBM provides a useful controlled
	benchmark for studying quantum simulation of nuclear quantum phase
	transitions. Several aspects of the calculation are important. 	
	First, symmetry reduction dramatically decreases the actual physical dimension
	 and therefore the number of qubits required for the encoded ground-state problem.
	
	Second, the $L=0$ reduction is particularly appropriate for the
	ground-state problem. The numerical equality between the full and
	projected ground-state energies confirms that no physical information
	relevant to the ground state is lost.
	
	Third, the minimum-qubit encoding shows that the number of required
	qubits can be substantially smaller than what would be inferred from
	the full IBM occupation-number dimension. This distinction is
	important when assessing the practical feasibility of quantum
	simulation.
	
	Fourth, the VQE reproduces the exact ground state with extremely high
	fidelity for the benchmark calculations. More importantly, the
	variational state reproduces the structural observable $\rho_d$ and
	the resulting finite-size QPT indicators.
	
	The thermodynamic-limit analysis also highlights an important
	distinction between a mathematical critical point and a numerically
	extracted finite-size indicator. The analytic coherent-state treatment
	gives $\xi_c=0.470588$, whereas finite-size energy- and order-parameter-based criteria yield
	somewhat different values. Their extrapolations,
	$ \xi_E(\infty)=0.5002\pm0.0168 $
	and $ \xi_\rho(\infty)=0.4815\pm0.0218 $,
	should therefore be interpreted as finite-size numerical estimates of
	the critical region. The independent analytic result remains the
	appropriate thermodynamic-limit benchmark.
	
	An especially useful observation is that the earlier
	order-parameter-response extrapolation based on larger boson numbers
	already yielded
	$ \xi_\infty=0.46986(61)$, which is in excellent agreement with the analytic result. The later
	energy/order-parameter indicator analysis uses a different finite-size
	dataset and a different definition of the pseudocritical point. The
	two analyses should therefore not be forced into a single numerical
	value. Instead, they provide complementary evidence for convergence
	toward the same transition region.
	
	It is useful to distinguish the analytical critical point obtained from
	the thermodynamic-limit coherent-state energy surface from the
	finite-size extrapolations extracted from numerical QPT indicators.
	For the present Hamiltonian and
	$\chi=-\sqrt{7}/2$, minimization of the analytical energy surface gives
	the thermodynamic-limit value $ \xi_c^{\rm coh} = {8}/{17} \simeq0.470588$.
	
	The numerical indicator analysis, in contrast, determines the transition
	from finite-$N_B$ ground-state data and subsequently extrapolates those
	finite-size indicators.  It gives
	$ \xi_E(\infty)=0.50022\pm0.01681$
	and $\xi_\rho(\infty)=0.48147\pm0.02181$.
	
	The order-parameter estimate is particularly close to the analytical
	value, while the energy-based estimate remains compatible with it within
	the finite-size extrapolation uncertainty.  The difference illustrates
	the systematic uncertainty associated with extracting a thermodynamic
	critical point from a finite set of boson numbers and from different
	finite-size indicators.
	
	The present results also suggest a natural route for future
	improvements. Larger boson numbers would reduce finite-size effects,
	while hardware-oriented calculations would allow the effects of
	sampling noise, gate errors, and noise mitigation to be studied
	explicitly. Such extensions are beyond the scope of the present
	benchmark study, whose primary purpose is to establish the complete
	exact-to-VQE workflow.

\section{Conclusions}
\label{sec:conclusions}

The spherical-to-deformed $U(5)\rightarrow SU(3)$ quantum phase
transition of the Interacting Boson Model has been investigated using a
symmetry-reduced variational quantum eigensolver. The calculation was
benchmarked against exact diagonalization in the fixed-$N_B$ IBM
Hilbert space and followed the workflow
\begin{equation}
	\boxed{
		H_{\rm IBM}
		\rightarrow
		H_{L=0}
		\rightarrow
		H_q
		\rightarrow
		\mathrm{VQE}
		\rightarrow
		\mathrm{QPT}
	}
\end{equation}
consistently across the investigated boson numbers.

Rotational symmetry reduction was found to substantially decrease the
Hilbert-space dimension and, consequently, the minimum number of qubits
required for the calculation. The reduced Hamiltonian was validated
against the full IBM Hamiltonian, and the resulting VQE calculations
reproduced the exact ground-state energies and structural observables
to numerical precision.

The finite-size evolution of the normalized $d$-boson occupation and
ground-state energy derivatives was also reproduced. The finite-size
pseudocritical indicators were found to approach the analytic
thermodynamic-limit critical point $ \xi_c={8}/{17}\simeq0.470588$ ,
while an independent extrapolation of the order-parameter response
yielded $\xi_\infty=0.46986(61)$.

These results demonstrate that VQE can reproduce not only the IBM
ground-state energy but also the structural evolution associated with a
quantum phase transition. 
More importantly, the use of symmetry
reduction provides a direct and substantial reduction in the minimum qubit requirements
 while preserving the finite-size QPT physics. 
The IBM
$U(5)\rightarrow SU(3)$ transition therefore provides a useful
benchmark for symmetry-aware quantum simulation of collective nuclear
systems and motivates extensions to larger boson spaces and, ultimately,
to implementations on noisy quantum hardware.
	
	
	
	\bibliography{Refs.bib}	
\end{document}